\documentclass[aps,pra,twocolumn,longbibliography,superscriptaddress,floatfix]{revtex4-1}
\usepackage{amsfonts}
\usepackage{mathtools}
\usepackage{graphicx}
\usepackage{epsfig}
\usepackage{dcolumn}% Align table columns on decimal point
\usepackage{bm}% bold math
\usepackage{amsmath}
\usepackage{graphicx}
\usepackage[latin1]{inputenc}
\usepackage{ulem}
\usepackage{epstopdf}
\usepackage{subfigure}
\usepackage{color}
\usepackage{amsthm}
\usepackage{newlfont}
\usepackage{graphicx}
\usepackage{epstopdf}
\usepackage{appendix}
\usepackage[breaklinks=true]{hyperref}
\usepackage{breakcites}
\usepackage{textcomp}
\usepackage{appendix}
\usepackage{multirow}	
\usepackage{color}
\usepackage{amssymb}
\usepackage{epsfig}
\usepackage{bm}
\usepackage[american]{babel}
\usepackage{braket}
\usepackage{soul}
\usepackage{float}
\hypersetup{colorlinks=true,linkcolor=blue,citecolor=blue,filecolor=blue,urlcolor=blue,pdfstartview=FitH}

\begin{document}
\title{
Dynamical Glass and Ergodization Times in Classical Josephson Junction Chains
}
\author{Thudiyangal Mithun}
 \affiliation{Center for Theoretical Physics of Complex Systems, Institute for Basic Science, Daejeon 34051, Korea}   
\author{Carlo Danieli}%
\affiliation{Center for Theoretical Physics of Complex Systems, Institute for Basic Science, Daejeon 34051, Korea}
\author{Yagmur Kati}%
    \affiliation{Center for Theoretical Physics of Complex Systems, Institute for Basic Science, Daejeon 34051, Korea}
\affiliation{Basic Science Program, Korea University of Science and Technology (UST), Daejeon 34113, Republic of Korea}
\author{Sergej Flach}%
%\email{ }%
\affiliation{Center for Theoretical Physics of Complex Systems, Institute for Basic Science, Daejeon 34051, Korea}

%%%%%%%%%%%%%%%%%%%%%%%%%%%%%%%%%%%%%%%%%%%
%%%%%%%%%%%%%%%             ABSTRACT                 %%%%%%%%%%%%%%
%%%%%%%%%%%%%%%%%%%%%%%%%%%%%%%%%%%%%%%%%%%
\begin{abstract}

Models of  classical Josephson junction chains turn integrable in the limit of large energy densities or small Josephson energies.
Close to these limits the Josephson coupling between the superconducting grains induces a short range nonintegrable network.
We compute distributions of finite time averages of grain charges and extract the ergodization time $T_E$ which controls their convergence to ergodic $\delta$-distributions. 
We relate $T_E$ to the statistics of fluctuation times of the charges, which are dominated by fat tails. 
$T_E$ is growing  anomalously fast upon approaching the integrable limit, as compared to the Lyapunov time $T_{\Lambda}$ - 
the inverse of the largest Lyapunov exponent - reaching astonishing ratios $T_E/T_{\Lambda} \geq 10^8$.
The microscopic reason for the observed dynamical glass is routed
in a growing number of grains evolving over long times in a regular almost integrable fashion due to the low probability of resonant interactions with the nearest neighbors.
We conjecture that the observed dynamical glass is a generic property of Josephson junction networks irrespective of their space dimensionality.

\end{abstract}

\maketitle 

Ergodicity is a core concept of statistical physics of many body systems. It demands infinite time averages of observables during a microcanonical evolution to match with their proper phase space averages \cite{lichtenberg1992regular}. Any laboratory or computational experiment is however constrained to finite averaging times. Are these sufficient or not? 
How much time is needed for a trajectory to visit the majority of the available microcanonical states, and for the finite-time average of an observable to be reasonably close to its statistical average? 
Can we define an {\it ergodization time} scale $T_E$ on which these properties manifest? What is that ergodization time depending on?
Doubts on the applicability of the ergodic hypothesis itself  were discussed for such simple cases as a mole of Ne at room temperature 
(see \cite{Gaveau2015ergodicity} and references therein).
Glassy dynamics have been reported in a large variety of Hamiltonian systems 
 \cite{Biroli:2017,Perez:2018,Tong:2018, Senanian:2018}.
Further, spin-glasses \cite{Bouchaud:1992} and stochastic Levy processes \cite{Bel:2005,Bel:2006,Rebenshtok:2007,Rebenshtok:2008,Korabel:2009,Schulz:2015}  reveal that the ergodization time (and even ergodicity itself) may be affected by heavy-tailed distributions of  lifetimes of typical excitations. 
The aim of this work is to address the above issues using a simple and paradigmatic dynamical many-body system testbed. 

Josephson junction networks are devices that are known for their wide applicability over various fields such as superconductivity, cold atoms, optics and meta-materials, among others \cite{Cataliotti:2001,Ryu:2013,Cassidy:2017} (for a recent survey on experimental results, see \cite{Blackburn:2016}): 
synchronization has been studied in Ref.  \cite{Tsygankov:2002,Nashar:2003}, discrete breathers were observed and studied in Ref. \cite{Binder:2000,Miroshnichenko:2001,Fistul:2002,Miroshnichenko:2005}, qubit dynamics
was analyzed in Ref. \cite{Shulga:2018,Martinis:2004} and the thermal conductivity was computed in Ref. \cite{Gendelman:2000,Giardina:2000}. 
In particular, a recent study conducted by Pino {\it et.al.}  \cite{Pino:2016} showed the existence of a non-ergodic/bad metal region in the high-temperature regime of a quantum chain of Josephson junctions, that exists as a prelude to a many-body localization phase \cite{Basko2006metal}. 
Notably, in \cite{Pino:2016} it has been conjectured that the bad metal regime persists as a non-ergodic phase in the classical limit of the model -
the large energy density regime of a chain of coupled rotors, close to an integrable limit. 
A similar prediction of a nonergodic phase (called weak coupling phase) in the same model was obtained in \cite{Escande:1994}. Further in \cite{Roeck:2014}, a faster decay of thermal conductivity in the high-temperature regime is observed. 
 On the other side, a strong slowing down of relaxations has been identified 
%a novel dynamical glass phase has been recently identified 
in the proximity of such a limit if the nonintegrable perturbation spans a short range network between corresponding actions \cite{Mithun:2018}. 
%This dynamical glass phase is still ergodic, though being characterized by rapidly increasing ergodization time
%scales.
The limit of weak Josephson coupling or high temperature is precisely corresponding to that short range network case. 
Is the Josephson junction chain then ergodic or not?

In this letter we demonstrate the existence of a dynamical glass in a classical Josephson junction chain of coupled rotors.  
We evaluate the 
convergence of distributions of finite time averages of the superconducting grain charges, and extract an ergodization time scale $T_E$. We show that this time scale is related to
the properties of the statistics of charge fluctuation times. Such fluctuation event statistics was introduced in Refs. \cite{Danieli:2017,Mithun:2018}. 
We compute the Lyapunov time $T_{\Lambda}$ - the inverse of the largest Lyapunov exponent $\Lambda$  \cite{Casetti:1996,Casetti:2000}. The Lyapunov time is a lower bound for the ergodization time: $T_{\Lambda} \leq T_E$. 
In the reported dynamical glass, the dynamics stays ergodic and $T_E$ is finite.
However, $T_E$ is growing  anomalously fast upon approaching the integrable limit, as compared to $T_{\Lambda}$ reaching astonishing ratios $T_E/T_{\Lambda} \geq 10^8$.
We show that $T_E$ is controlled by fat tails of charge fluctuation time distributions. We compute the spatio-temporal evolution of nonlinear resonances between interacting grains \cite{Livi:1987,Escande:1994}.
The microscopic reason for the observed dynamical glass is routed
in a growing number of grains evolving over long times in a regular almost integrable fashion due to the low probability of resonant interactions with nearest neighbors.
The dynamical glass behavior is expected to be a generic property of a large class of dynamical systems,
where ergodization time scales depend sensitively on control parameters. At the same time, the concept of ergodicity is preserved, and statistical physics continues to work - it is all just a matter of time scales.

We consider the Hamiltonian
\begin{equation}
\small  H(q,p) =\sum_{n=1}^{N}\bigg[\frac{p_n^2}{2}+E_J(1-\cos(q_{n+1}-q_n)) \bigg], 
\label{eq:rotor1}
\end{equation}
describing the dynamics of a chain of $N$ superconducting islands with weak nearest neighbor Josephson coupling in its classical limit.
We note that this model is equivalent to an XY chain or similarly to a coupled rotor chain, where the grain charging energies turn into the above kinetic energy terms \cite{Livi:1987, Pino:2016}.
We apply periodic boundary conditions $p_1 = p_{N+1}$ and $q_{1} = q_{N+1}$ for the conjugate angles $q_n$ and momenta $p_n$. $E_J$ controls the strength of Josephson coupling.
The corresponding equations of motion of Eq.~(\ref{eq:rotor1}) are 
\begin{equation}
  \begin{split}
\small  \dot{q_n} =  p_n ,~~~~\dot{p_n}  
& = E_J \big[  \sin(q_{n+1}-q_n)+\sin(q_{n-1} - q_n)  \big].
\end{split}
\label{eq:rotor2}
\end{equation}
This system has two conserved quantities: the total energy $H$ and the total angular momentum $L=\sum_{n=1}^N p_n$. 
We will choose $L=0$ without loss of generality. Exact expressions for average full $h$ and kinetic $k$ energy densities as functions of temperature are obtained using a Gibbs distribution \cite{Supple} and yield
\begin{equation}
h = k +E_J\left( 1 - \frac{I_1(E_J/2k)}{I_0(E_J/2k)} \right) \;,
\label{h(k)}
\end{equation}
with $I_{0,1}$ being the modified Bessel functions of the first kind.
We investigate the equilibrium dynamics of the above system in proximity to two integrable limits: $h \rightarrow \infty$, or $E_J \rightarrow 0$.
At these limits, the system reduces to a set of uncoupled superconducting grains $H_0 = \sum_{n=1}^{N}\frac{p_n^2}{2}$ \cite{Livi:1987}. 
In proximity to these limits the Josephson terms induce a nonintegrable perturbation through a short-range interaction network of actions $\{p_n\}_n$ \cite{Mithun:2018}.  
We consider the kinetic energies $k_n = p_n^2 / 2$ as a set of time-dependent observables. Due to the discrete translational invariance of $H$ all $k_n$ variables are
statistically equivalent, fluctuating around their equilibrium value $k$.   
We will integrate the equations of motion using symplectic integrators \cite{Supple}. Unless, otherwise stated, we use the system size $N=2^{10}$.

%--------------------------------------------                
  \begin{figure}[!htbp] 
    \includegraphics[scale=1.0,width=0.48\textwidth]{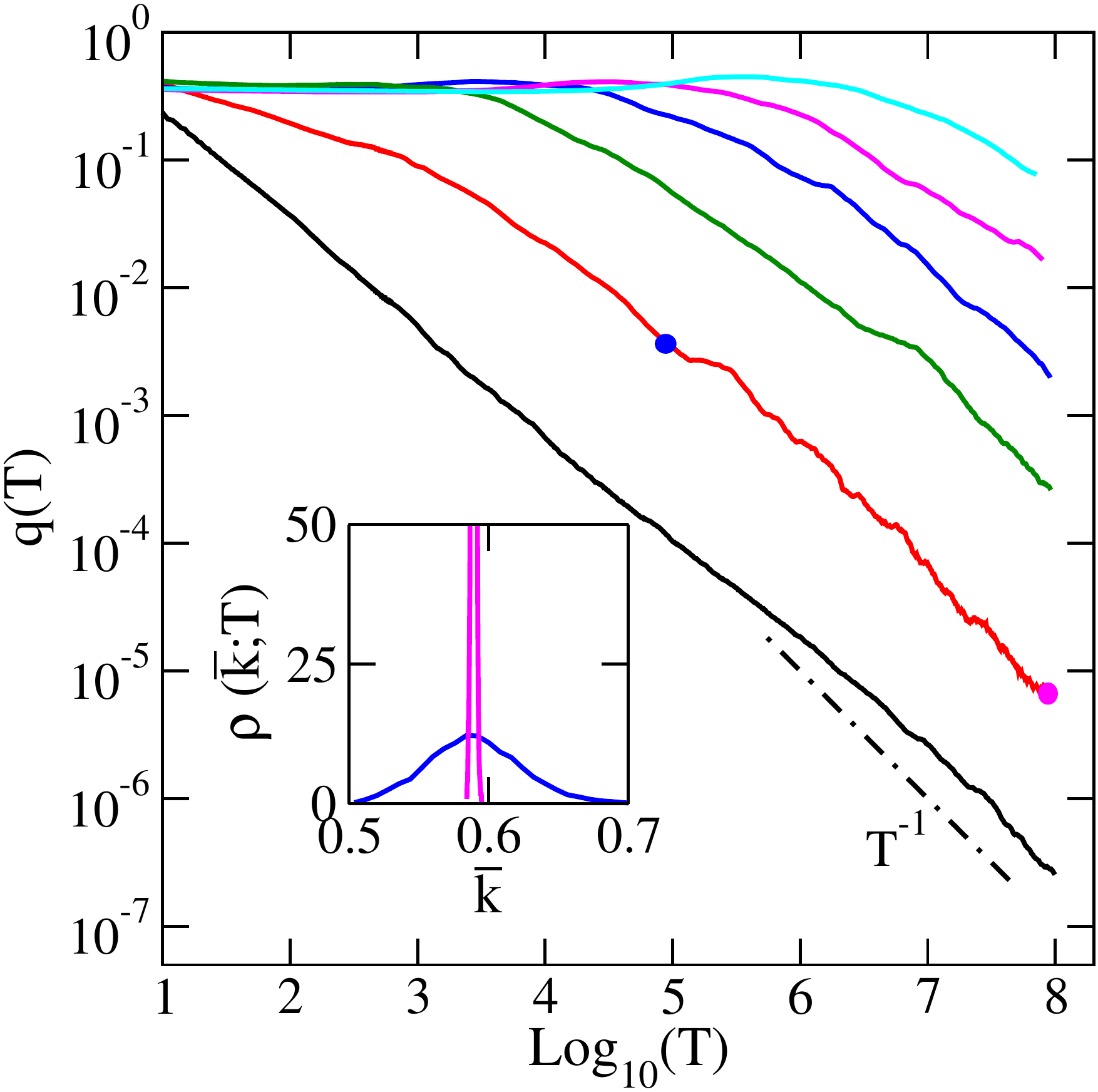}\\ %\vspace{-0.8em}
  \caption{\label{fig1}{\footnotesize (Color online) a) Fluctuation Index $q(T)$ for energy densities  (bottom to top) 0.1 (black) 1.2 (red), 2.4 (green), 3.8 (blue), 5.4 (magenta), and 8.5 (cyan). Inset: $\rho(\overline{k};T)$ for $h=1.2$ and two different times $T=10^5$ (blue circle)  and $T=10^8$ (magenta circle) marked in the main plot.  
Here $E_J=1$ and  the number of realizations $R=12$ for all data.
}
}
   \end{figure}
%--------------------------------------------
      
To quantitatively assess the ergodization time $T_E$, we compute finite time averages $\overline{k}_{n,T} = \frac{1}{T} \int_0^{T} k_n(t)dt$ 
for a set of $R$ different trajectories at given $h,E_J$. The corresponding distribution $\rho(\overline{k};T)$ is characterized by its 1st moment $\mu_k(T)$ and the standard deviation $\sigma_k(T)$.
Assuming ergodicity, $\mu_k(T \rightarrow \infty) =k$ and $\sigma_k(T \rightarrow \infty) =0$, since the distribution $\rho(\overline{k};T \rightarrow \infty) = \delta(\overline{k}-k)$. 
In the inset of Fig.~\ref{fig1} we show the distributions $\rho(\overline{k};T)$ for $h=1.2$  at two different averaging times $T=10^5,10^8$.
As expected the distribution $\rho(\overline{k},T)$ converges to a delta function, centered around $k$. 
We then use 
the fluctuation index $q(T)=\frac{\sigma_k^2(T) }{\mu_k^2(T)}$ as a quantitative dimensionless measure of the above convergence properties.

In Fig.~\ref{fig1} we show $q(T)$ for different values of $h$ with $E_J=1$. We find $q(T \ll T_E) = q(0)$ and $q(T \gg T_E) \sim T_E/T$ where $T_E$ is our definition of the ergodization time scale.
We rescale and fit the different curves $q(T)$ and extract $T_E$ \cite{Supple}. The result is plotted in Fig.~\ref{fig3}(a) with green squares.
$T_E$ quickly grows by orders of magnitude upon increasing the energy density $h$ in a rather moderate window of values, 
close to a power law $T_E \sim h^6$.
When fixing $h=1$ and varying $E_J$, we make similar observations  \cite{Supple}, with $T_E \sim E_J^{-6.5}$ as shown in  Fig.~\ref{fig3}(b). With our results, we validate ergodic dynamics in the considered system. Previous
reports \cite{Pino:2016,Escande:1994} were not addressing the quickly growing time scale $T_E$ upon 
approaching the integrable limit.
%
%--------------------------------------------
     \begin{figure}[!htbp] 
    \includegraphics[scale=1,width=0.48\textwidth]{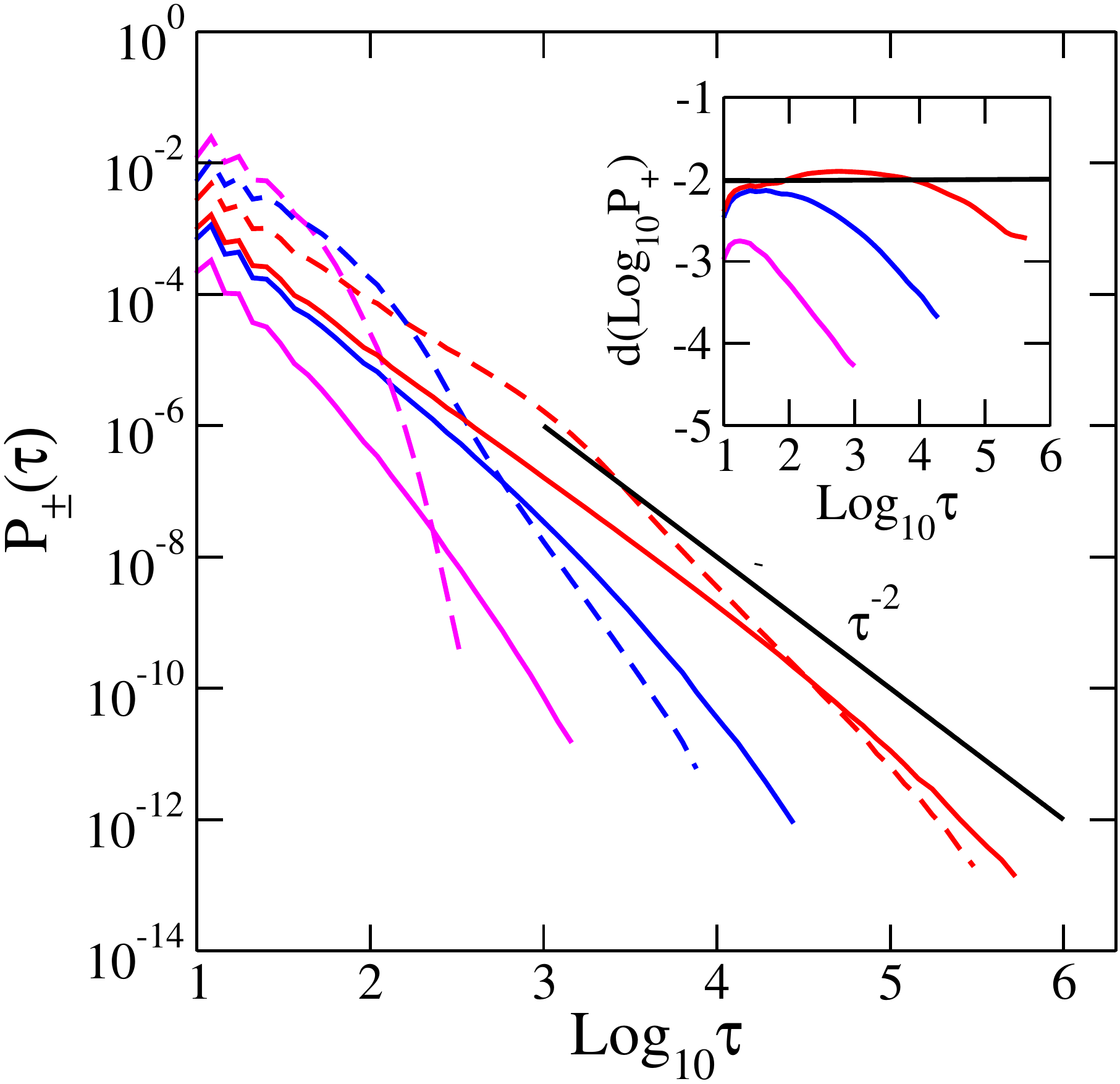}
  \caption{\label{fig2}{\footnotesize (Color online) $P_{+}(\tau)$ (solid lines) and $P_{-}(\tau)$ (dashed lines) for  various energy densities 
$h=1.2$ (magenta), $h=2.4$ (blue) and $h=5.4$ (red). Here $E_J=1$.  
The black line corresponds to a power law decay $\tau^{-2}$ and guides the eye.
Inset:  $\gamma \equiv d(\log_{10} P_+)/d(\log_{10} \tau)$. 
The black solid horizontal line guides the eye at value -2.
 }}
   \end{figure}
   %-------------------------------------------
   
Let us study the fluctuation statistics of the observables $k_n(t)$. Each of them has to fluctuate around their common average $k$. This allows to
segment the trajectory of the whole system phase space into consecutive excursions \cite{Danieli:2017,Mithun:2018}. Note that for each site $n$ the
segmenting is different, and we account for all of them. 
We measure the consecutive piercing times $t_n^{i}$ at which $k_n(t)=k$. 
We then compute the {\it excursion times} $\tau^{\pm}_n(i) = t_n^{i+1} -t_n^i$ for a trajectory of excursion events during which $k(t) > k$ ($\tau^+$) and  $k(t) < k$ ($\tau^-$) respectively.
Fig.~\ref{fig2} shows the distributions for $E_J=1$ and various energy densities $h$.  
As $h$ increases both distributions increase  their tail weights, with $P_{+}$ dominating over $P_{-}$. Further, the distributions develop 
intermediate tail structures close to a $1/\tau^2$. (inset in Fig.\ref{fig2}). 

We can now compute the following two time scales:  the average excursion time, $\mu_\tau$ and the standard deviation $\sigma_\tau$ of the distribution $P_+$, which are shown in Fig.\ref{fig3}(a) (orange diamonds and blue triangles) as functions of $h$. We observe that $\sigma_{\tau}$
equals with $\mu_{\tau}$ at $h \approx 1$ and quickly overgrows $\mu_{\tau}$ for $h > 1$, signaling the 
proximity to an integrable limit,  where the dynamics is dominated by fluctuations rather than the means.
Indeed, if the distributions $P_{\pm}({\tau})$ asymptotically reach a $1/\tau^2$ dependence in the integrable limit,
then not only the time scales $\mu_{\tau}$ and $\sigma_{\tau}$ have to diverge, but their ratio 
$\sigma_{\tau}/\mu_{\tau}$ will diverge as well.

The above time scales are related to the ergodization time
scale $T_E$ as
\begin{equation}
T_E \sim \tau_q \equiv \frac{\sigma^2_{\tau}}{\mu_{\tau}}. 
\label{popcorn}
\end{equation}
This relation can be obtained e.g. after approximating the dependence $k_n(t)$ by a telegraph random process
with excursion time distributions $P_{\pm}$ \cite{Supple}. We plot $A \tau_q$ versus $h$ in Fig.\ref{fig3}(a)
(black circles) with a fitting parameter $A=130$. The curve is strikingly close to the dependence $T_E(h)$
for $h > 1$. We thus independently reconfirm that the considered system dynamics is ergodic, yet with 
quickly growing time scales of ergodization.
When fixing $h=1$ and varying $E_J$, we make similar observations as shown in  Fig.~\ref{fig3}(b).
 The physical origin of $A$ is an interesting and open question for future studies. 

 %------------------------------
                  \begin{figure}[!htbp] 
     \includegraphics[width=8.5cm,scale=1]{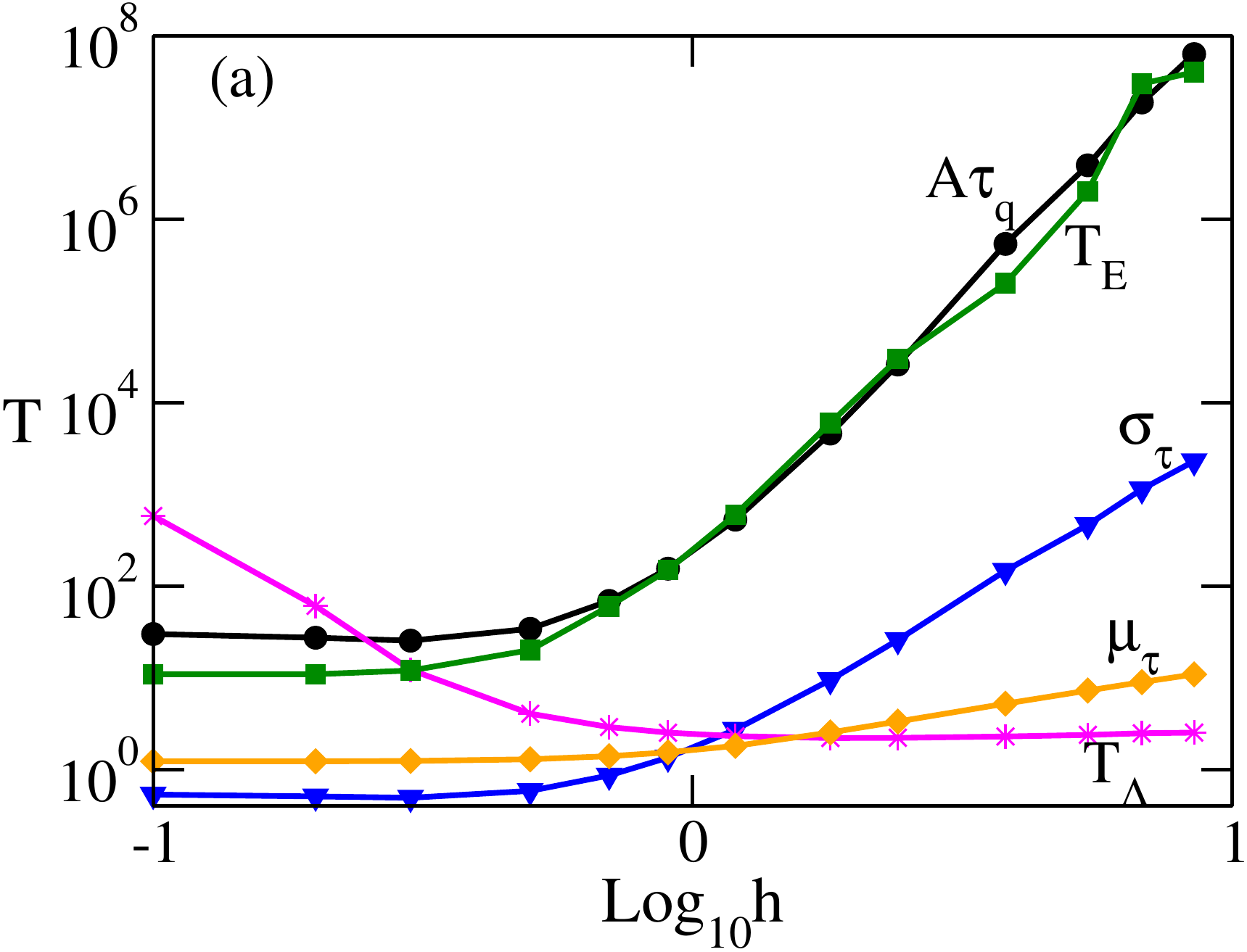}
    \includegraphics[width=8.5cm,scale=1]{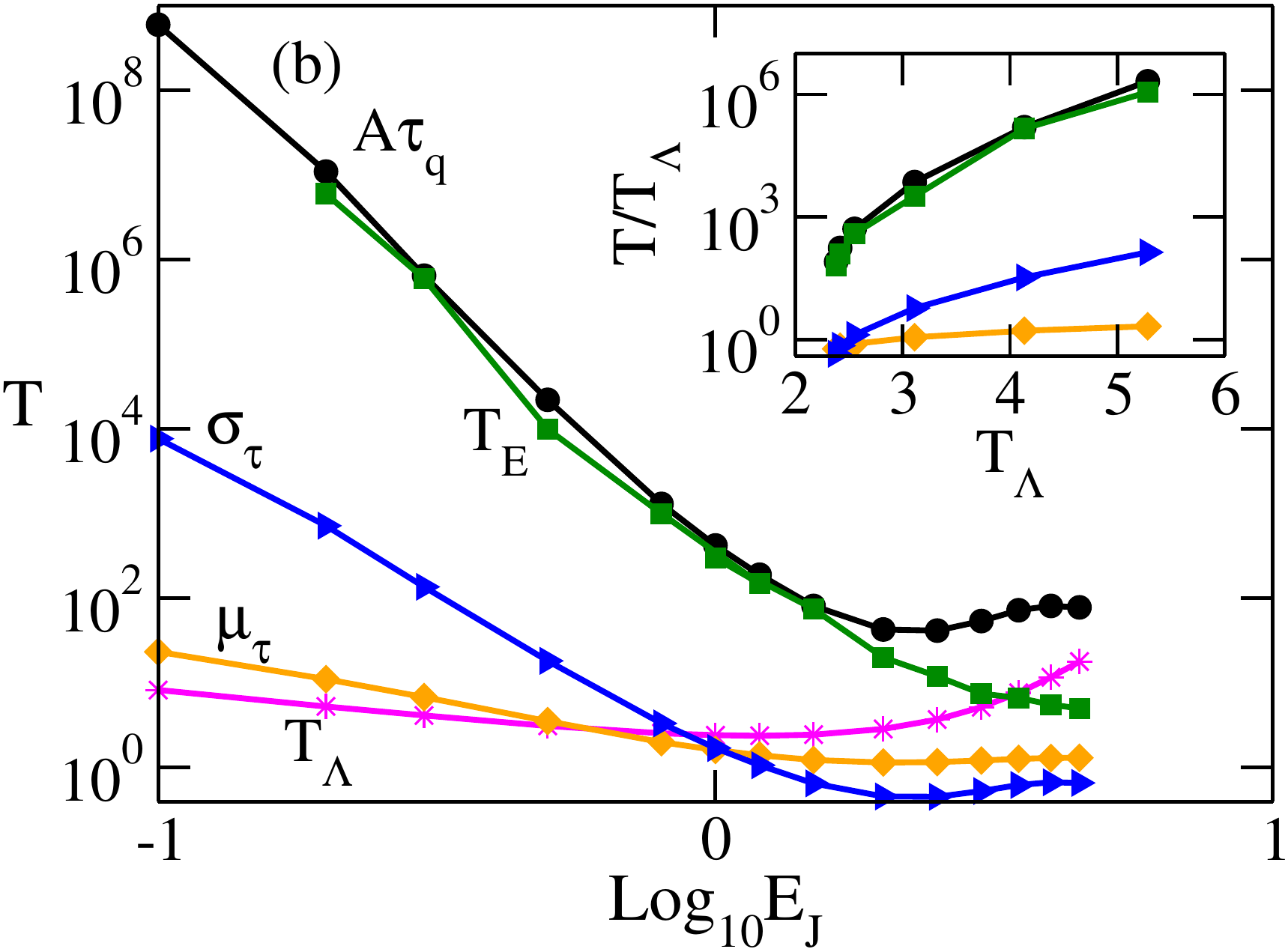} 
  \caption{\label{fig3}{\footnotesize (Color online) 
(a)
Time scales  $T_E$ (green squares), $A \tau_q$ (black circles), Lyapunov time $T_{\Lambda}$ (magenta stars), $\mu_{\tau}$ (orange diamonds) and $\sigma_{\tau}$ (blue triangles)
  vs the energy density $h$ for $E_J=1$ and $A=130$.
(b) Same as in (a) but vs
  $E_J$ at the fixed energy density $h=1$. 
Inset:  Rescaled times ($A \tau_q$, $T_E$, $m_{\tau}$  and $\sigma_{\tau}$) in units of the Lyapunov time $T_{\Lambda}$.
  }}
   \end{figure}
   %--------------------------------
        \begin{figure}[!htbp] 
                              \includegraphics[width=8.5cm,scale=1]{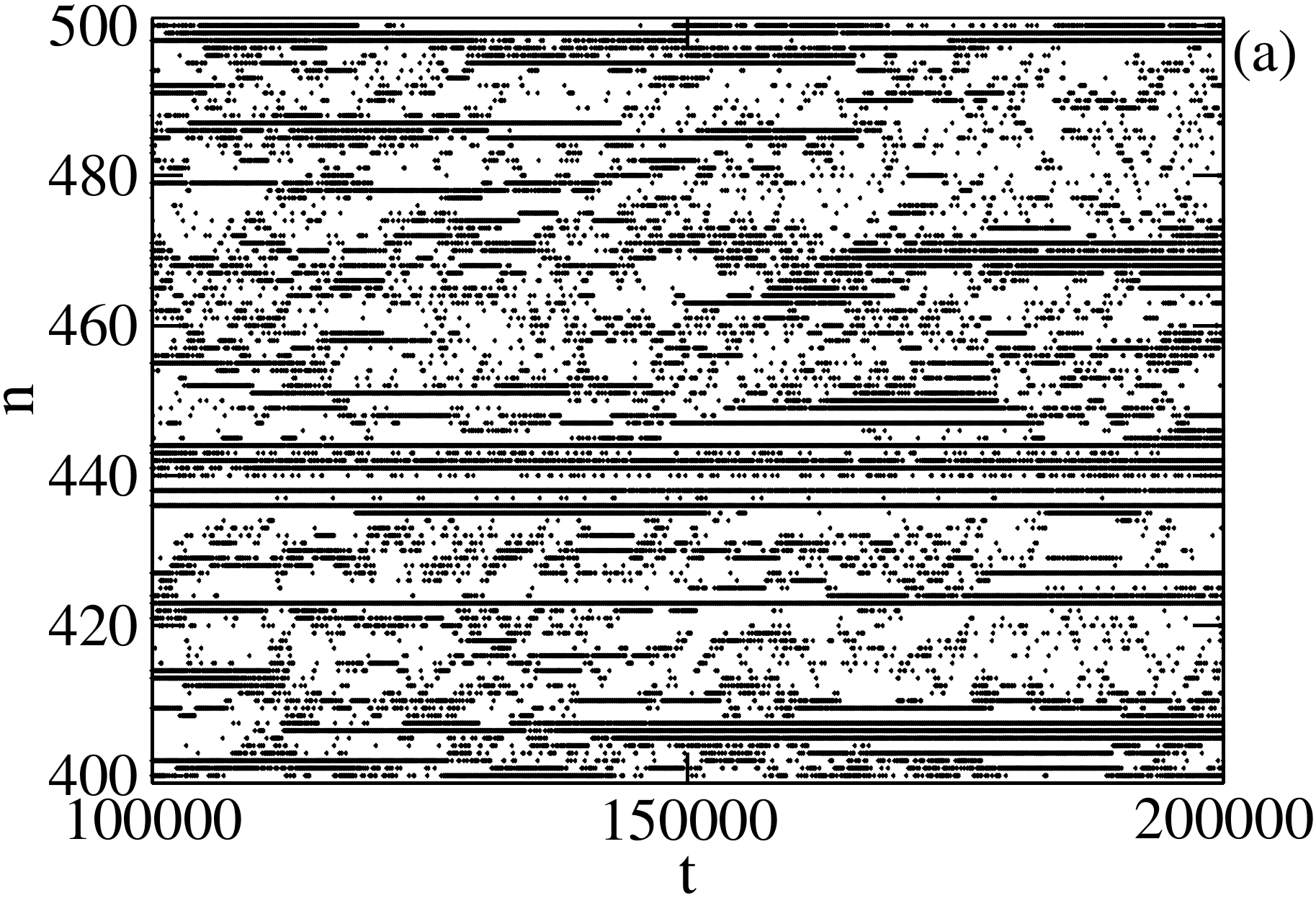} \\
                              \includegraphics[width=8.5cm,scale=1]{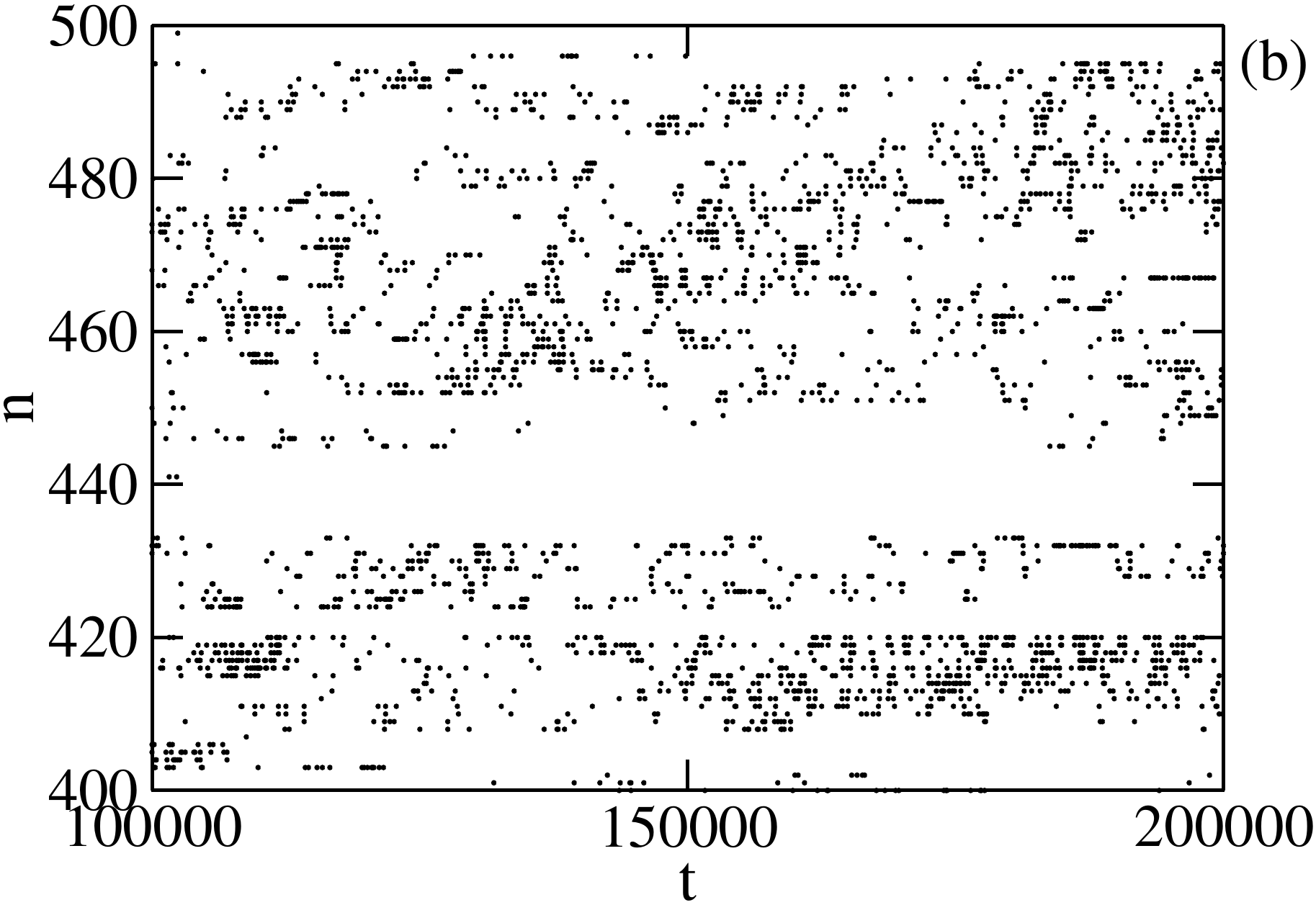} 
  \caption{\label{fig4}{\footnotesize (Color online) (a)  Spatiotemporal evolution of a part of the Josephson junction chain for $E_J=1$ and $h=5.4$.
Black points correspond to events with $k_n > k$. 
(b) Spatiotemporal evolution of nonlinear resonances for the same parameters as in (a). See text for details.
  } }
   \end{figure}
      %--------------------------------   

With ergodicity being restored, the question remains: what is the microscopic origin of the enormously fast
growing ergodization time? Since the considered system is nonintegrable, its dynamics must be chaotic.
Therefore there is a Lyapunov time scale $T_{\Lambda}= 1/\Lambda$ dictated by the largest Lyapunov exponent $\Lambda$. $T_{\Lambda}$ can be expected to serve as a lower bound for the ergodization time scale.
We compute $\Lambda$ using standard techniques \cite{Supple} and also compare it with theoretical predictions   \cite{Casetti:1996,Casetti:2000}. 
We plot the Lyapunov time $T_{\Lambda}$ versus $h$ in Fig.~\ref{fig3}(a) and versus $E_J$ in Fig.~\ref{fig3}(b) (magenta stars).
The surprising finding is that $T_{\Lambda} \lesssim \mu_{\tau}$ in proximity to the integrable limit. Thus 
$T_{\Lambda} \ll T_E$, e.g. for $E_J=1$ and $h=10$ we find $T_{\Lambda} \sim 1$ and $T_E \sim 10^8$. 
The inset of Fig.~\ref{fig3}(b) demonstrates the above findings where $T_E$, $A\tau_q$, $\mu_{\tau}$ and $\sigma_{\tau}$ are 
plotted versus $T_{\Lambda}$ and in units of $T_{\Lambda}$.  
These are typical features of the  novel dynamical glass,  
%introduced in \cite{Danieli:2018}, 
which
starts at $h\approx E_J$.

In order to advance, we analyze the spatiotemporal dynamics of $k_n(t)$ in Fig.~\ref{fig4}(a).
We plot black points during events $k_n(t) > k$ for $E_J=1$ and $h=5.4$ over a  time window of $10^5$ and a spatial window of  100 sites.
We observe many long lasting events, which slowly diffuse in space. At the same time, regions between these long lasting events appear to be more
chaotic, with this chaos however being confined to regions between two events. The events correspond
to long-living breather-like excitations \cite{Mithun:2018}. 
The existence of chaotic spots was predicted in Refs. \cite{Pino:2016,Escande:1994}. We then compute
the  frequency difference $\Delta_n = |\omega_n-\omega_{n+1}|$ between neighboring grains, where $\omega_n = \dot{q}_n = p_n$. 
Following \cite{Chirikov:1979,Pino:2016,Escande:1994},  we define a chaotic resonance if a neighbouring pair $\Delta_n <1$ and $\Delta_{n+1}<1$.
We plot the spatiotemporal evolution of these nonlinear resonances in Fig.~\ref{fig4}(b) with the resonances marked with black dots. We observe a slowly diffusing and meandering network of chaotic puddles.

The large ergodization time $T_E$ could be related to a small density of chaotic spots, and/or to a weak interaction between the spots.
The density of chaotic spots was calculated in \cite{Escande:1994} as 
$D= \frac{1}{\sqrt{\pi}}\int_0^ye^{-x^2}dx$
where $y=\sqrt{\frac{16\beta}{3}} $ for $E_J=1$ and $\beta$ is the inverse temperature \cite{Supple}. Note that $1/\beta \approx  2 h$ for $h \gg 1$.
It follows that $D \sim 1/\sqrt{h}$ for $h \gg 1$. This decay is way too slow to explain the rapid increase of the ergodization time $T_E$ upon increasing $h$ in
Fig.~\ref{fig3}(a). There $h$ increases by one order of magnitude, $D$ decreases by a factor of 3, but $T_E$ increases by six orders of magnitude.
Therefore the ergodization time in the dynamical glass must be controlled by a very weak interaction between chaotic spots, which have to
penetrate silent non-chaotic regions formed by breather like events.

To conclude, the classical dynamics of a Josephson junction chain at large temperatures (i.e. energy densities) or likewise at weak Josephson coupling
is characterized by a dynamical glass in its proximity to corresponding integrable limits. This dynamical glass is induced by the short range of the nonintegrable perturbation network
spanned between the actions which turn integrals of motion at the very integrable limit. The dynamics of the system remains ergodic, albeit with
rapidly increasing ergodization time $T_E$. We relate $T_E$ to time scales extracted from the fluctuations of the actions. We also show, that
the Lyapunov time, which is marking the onset of chaos in the system, is orders of magnitude shorter than $T_E$. The reason for the rapidly growing
ergodization time is rooted in the slowing down of interactions between chaotic spots. By virtue of the short range network we expect our results
to hold as well in higher space dimensions. 
A highly nontrivial and interesting question is the relation of the dynamical glass to the KAM regime  \cite{Kolmogorov1954conservation, arnold1963aproof,Moser1962invariant}.
Common expectations tell that the KAM regime thresholds of a nonintegrable perturbation vanish very fast with an increasing number of degrees of freedom $N$, perhaps
even exponentially fast due to proliferating resonances
 \cite{Chierchia1982smooth,Benettin1984proof,Wayne1984KAM1,Wayne1984KAM2}. The long-lasting regular motion in the dynamical glass is local both in space and
time, as manifested by the exponential cutoff tails in Fig.\ref{fig2}. The dynamical glass appears to have support from a chaotic component of measure one in the 
available phase space. This is very different from few degree of freedom systems with a mixed phase space. 
A quantitative theory for the dependence of the ergodization time on the control parameters in the proximity to
the discussed integrable limits is a challenging future task, and as intriguing as the question about the fate of the dynamical glass in the
related quantum many body problem.

The authors acknowledge financial support from IBS (Project Code No. IBS-R024-D1). We thank I. Vakulchyk, A. Andreanov, and M. Fistul for helpful discussions.

  \bibliographystyle{apsrev4}
\let\itshape\upshape
\normalem
\bibliography{reference1}

%merlin.mbs apsrev4-1.bst 2010-07-25 4.21a (PWD, AO, DPC) hacked
%Control: key (0)
%Control: author (72) initials jnrlst
%Control: editor formatted (1) identically to author
%Control: production of article title (1) required
%Control: page (0) single
%Control: year (1) truncated
%Control: production of eprint (0) enabled
\providecommand{\noopsort}[1]{}\providecommand{\singleletter}[1]{#1}%
\begin{thebibliography}{49}%
\makeatletter
\providecommand \@ifxundefined [1]{%
 \@ifx{#1\undefined}
}%
\providecommand \@ifnum [1]{%
 \ifnum #1\expandafter \@firstoftwo
 \else \expandafter \@secondoftwo
 \fi
}%
\providecommand \@ifx [1]{%
 \ifx #1\expandafter \@firstoftwo
 \else \expandafter \@secondoftwo
 \fi
}%
\providecommand \natexlab [1]{#1}%
\providecommand \enquote  [1]{``#1''}%
\providecommand \bibnamefont  [1]{#1}%
\providecommand \bibfnamefont [1]{#1}%
\providecommand \citenamefont [1]{#1}%
\providecommand \href@noop [0]{\@secondoftwo}%
\providecommand \href [0]{\begingroup \@sanitize@url \@href}%
\providecommand \@href[1]{\@@startlink{#1}\@@href}%
\providecommand \@@href[1]{\endgroup#1\@@endlink}%
\providecommand \@sanitize@url [0]{\catcode `\\12\catcode `\$12\catcode
  `\&12\catcode `\#12\catcode `\^12\catcode `\_12\catcode `\%12\relax}%
\providecommand \@@startlink[1]{}%
\providecommand \@@endlink[0]{}%
\providecommand \url  [0]{\begingroup\@sanitize@url \@url }%
\providecommand \@url [1]{\endgroup\@href {#1}{\urlprefix }}%
\providecommand \urlprefix  [0]{URL }%
\providecommand \Eprint [0]{\href }%
\providecommand \doibase [0]{http://dx.doi.org/}%
\providecommand \selectlanguage [0]{\@gobble}%
\providecommand \bibinfo  [0]{\@secondoftwo}%
\providecommand \bibfield  [0]{\@secondoftwo}%
\providecommand \translation [1]{[#1]}%
\providecommand \BibitemOpen [0]{}%
\providecommand \bibitemStop [0]{}%
\providecommand \bibitemNoStop [0]{.\EOS\space}%
\providecommand \EOS [0]{\spacefactor3000\relax}%
\providecommand \BibitemShut  [1]{\csname bibitem#1\endcsname}%
\let\auto@bib@innerbib\@empty
%</preamble>
\bibitem [{\citenamefont {Lichtenberg}\ and\ \citenamefont
  {Lieberman}(1992)}]{lichtenberg1992regular}%
  \BibitemOpen
  \bibfield  {author} {\bibinfo {author} {\bibfnamefont {A.~J.}\ \bibnamefont
  {Lichtenberg}}\ and\ \bibinfo {author} {\bibfnamefont {M.~A.}\ \bibnamefont
  {Lieberman}},\ }\bibfield  {title} {\enquote {\bibinfo {title} {Regular and
  {C}haotic {D}ynamics, vol. 38 of},}\ }\href@noop {} {\bibfield  {journal}
  {\bibinfo  {journal} {Applied Mathematical Sciences}\ } (\bibinfo {year}
  {1992})}\BibitemShut {NoStop}%
\bibitem [{\citenamefont {Gaveau}\ and\ \citenamefont
  {Schulman}(2015)}]{Gaveau2015ergodicity}%
  \BibitemOpen
  \bibfield  {author} {\bibinfo {author} {\bibfnamefont {B.}~\bibnamefont
  {Gaveau}}\ and\ \bibinfo {author} {\bibfnamefont {L.~S.}\ \bibnamefont
  {Schulman}},\ }\bibfield  {title} {\enquote {\bibinfo {title} {Is ergodicity
  a reasonable hypothesis for macroscopic systems?}}\ }\href {\doibase
  10.1140/epjst/e2015-02434-7} {\bibfield  {journal} {\bibinfo  {journal} {The
  European Physical Journal Special Topics}\ }\textbf {\bibinfo {volume}
  {224}},\ \bibinfo {pages} {891} (\bibinfo {year} {2015})}\BibitemShut
  {NoStop}%
\bibitem [{\citenamefont {Biroli}\ and\ \citenamefont
  {Tarzia}(2017)}]{Biroli:2017}%
  \BibitemOpen
  \bibfield  {author} {\bibinfo {author} {\bibfnamefont {G.}~\bibnamefont
  {Biroli}}\ and\ \bibinfo {author} {\bibfnamefont {M.}~\bibnamefont
  {Tarzia}},\ }\bibfield  {title} {\enquote {\bibinfo {title} {Delocalized
  glassy dynamics and many-body localization},}\ }\href {\doibase
  10.1103/PhysRevB.96.201114} {\bibfield  {journal} {\bibinfo  {journal} {Phys.
  Rev. B}\ }\textbf {\bibinfo {volume} {96}},\ \bibinfo {pages} {201114}
  (\bibinfo {year} {2017})}\BibitemShut {NoStop}%
\bibitem [{\citenamefont {P{\'e}rez-Espigares}\ \emph
  {et~al.}(2018)\citenamefont {P{\'e}rez-Espigares}, \citenamefont {Carollo},
  \citenamefont {Garrahan},\ and\ \citenamefont {Hurtado}}]{Perez:2018}%
  \BibitemOpen
  \bibfield  {author} {\bibinfo {author} {\bibfnamefont {C.}~\bibnamefont
  {P{\'e}rez-Espigares}}, \bibinfo {author} {\bibfnamefont {F.}~\bibnamefont
  {Carollo}}, \bibinfo {author} {\bibfnamefont {J.~P.}\ \bibnamefont
  {Garrahan}}, \ and\ \bibinfo {author} {\bibfnamefont {P.~I.}\ \bibnamefont
  {Hurtado}},\ }\bibfield  {title} {\enquote {\bibinfo {title} {Dynamical
  criticality in driven systems: non-perturbative results, microscopic origin
  and direct observation},}\ }\href@noop {} {\bibfield  {journal} {\bibinfo
  {journal} {arXiv preprint arXiv:1807.10235}\ } (\bibinfo {year}
  {2018})}\BibitemShut {NoStop}%
\bibitem [{\citenamefont {Tong}\ and\ \citenamefont
  {Tanaka}(2018)}]{Tong:2018}%
  \BibitemOpen
  \bibfield  {author} {\bibinfo {author} {\bibfnamefont {H.}~\bibnamefont
  {Tong}}\ and\ \bibinfo {author} {\bibfnamefont {H.}~\bibnamefont {Tanaka}},\
  }\bibfield  {title} {\enquote {\bibinfo {title} {Revealing {H}idden
  {S}tructural {O}rder {C}ontrolling {B}oth {F}ast and {S}low {G}lassy
  {D}ynamics in {S}upercooled {L}iquids},}\ }\href {\doibase
  10.1103/PhysRevX.8.011041} {\bibfield  {journal} {\bibinfo  {journal} {Phys.
  Rev. X}\ }\textbf {\bibinfo {volume} {8}},\ \bibinfo {pages} {011041}
  (\bibinfo {year} {2018})}\BibitemShut {NoStop}%
\bibitem [{\citenamefont {Senanian}\ and\ \citenamefont
  {Narayan}(2018)}]{Senanian:2018}%
  \BibitemOpen
  \bibfield  {author} {\bibinfo {author} {\bibfnamefont {A.}~\bibnamefont
  {Senanian}}\ and\ \bibinfo {author} {\bibfnamefont {O.}~\bibnamefont
  {Narayan}},\ }\bibfield  {title} {\enquote {\bibinfo {title} {Glassy dynamics
  in disordered oscillator chains},}\ }\href {\doibase
  10.1103/PhysRevE.97.062110} {\bibfield  {journal} {\bibinfo  {journal} {Phys.
  Rev. E}\ }\textbf {\bibinfo {volume} {97}},\ \bibinfo {pages} {062110}
  (\bibinfo {year} {2018})}\BibitemShut {NoStop}%
\bibitem [{\citenamefont {Bouchaud}(1992)}]{Bouchaud:1992}%
  \BibitemOpen
  \bibfield  {author} {\bibinfo {author} {\bibfnamefont {J.-P.}\ \bibnamefont
  {Bouchaud}},\ }\bibfield  {title} {\enquote {\bibinfo {title} {Weak
  ergodicity breaking and aging in disordered systems},}\ }\href@noop {}
  {\bibfield  {journal} {\bibinfo  {journal} {Journal de Physique I}\ }\textbf
  {\bibinfo {volume} {2}},\ \bibinfo {pages} {1705} (\bibinfo {year}
  {1992})}\BibitemShut {NoStop}%
\bibitem [{\citenamefont {Bel}\ and\ \citenamefont {Barkai}(2005)}]{Bel:2005}%
  \BibitemOpen
  \bibfield  {author} {\bibinfo {author} {\bibfnamefont {G.}~\bibnamefont
  {Bel}}\ and\ \bibinfo {author} {\bibfnamefont {E.}~\bibnamefont {Barkai}},\
  }\bibfield  {title} {\enquote {\bibinfo {title} {Weak ergodicity breaking in
  the continuous-time random walk},}\ }\href@noop {} {\bibfield  {journal}
  {\bibinfo  {journal} {Physical Review Letters}\ }\textbf {\bibinfo {volume}
  {94}},\ \bibinfo {pages} {240602} (\bibinfo {year} {2005})}\BibitemShut
  {NoStop}%
\bibitem [{\citenamefont {{Bel, G.}}\ and\ \citenamefont {{Barkai,
  E.}}(2006)}]{Bel:2006}%
  \BibitemOpen
  \bibfield  {author} {\bibinfo {author} {\bibnamefont {{Bel, G.}}}\ and\
  \bibinfo {author} {\bibnamefont {{Barkai, E.}}},\ }\bibfield  {title}
  {\enquote {\bibinfo {title} {Weak ergodicity breaking with deterministic
  dynamics},}\ }\href {\doibase 10.1209/epl/i2005-10501-8} {\bibfield
  {journal} {\bibinfo  {journal} {Europhys. Lett.}\ }\textbf {\bibinfo {volume}
  {74}},\ \bibinfo {pages} {15} (\bibinfo {year} {2006})}\BibitemShut {NoStop}%
\bibitem [{\citenamefont {Rebenshtok}\ and\ \citenamefont
  {Barkai}(2007)}]{Rebenshtok:2007}%
  \BibitemOpen
  \bibfield  {author} {\bibinfo {author} {\bibfnamefont {A.}~\bibnamefont
  {Rebenshtok}}\ and\ \bibinfo {author} {\bibfnamefont {E.}~\bibnamefont
  {Barkai}},\ }\bibfield  {title} {\enquote {\bibinfo {title} {Distribution of
  {T}ime-{A}veraged {O}bservables for {W}eak {E}rgodicity {B}reaking},}\ }\href
  {\doibase 10.1103/PhysRevLett.99.210601} {\bibfield  {journal} {\bibinfo
  {journal} {Phys. Rev. Lett.}\ }\textbf {\bibinfo {volume} {99}},\ \bibinfo
  {pages} {210601} (\bibinfo {year} {2007})}\BibitemShut {NoStop}%
\bibitem [{\citenamefont {Rebenshtok}\ and\ \citenamefont
  {Barkai}(2008)}]{Rebenshtok:2008}%
  \BibitemOpen
  \bibfield  {author} {\bibinfo {author} {\bibfnamefont {A.}~\bibnamefont
  {Rebenshtok}}\ and\ \bibinfo {author} {\bibfnamefont {E.}~\bibnamefont
  {Barkai}},\ }\bibfield  {title} {\enquote {\bibinfo {title} {Weakly
  {N}on-{E}rgodic {S}tatistical {P}hysics},}\ }\href {\doibase
  10.1007/s10955-008-9610-3} {\bibfield  {journal} {\bibinfo  {journal} {J.
  Stat. Phys.}\ }\textbf {\bibinfo {volume} {133}},\ \bibinfo {pages} {565}
  (\bibinfo {year} {2008})}\BibitemShut {NoStop}%
\bibitem [{\citenamefont {Korabel}\ and\ \citenamefont
  {Barkai}(2009)}]{Korabel:2009}%
  \BibitemOpen
  \bibfield  {author} {\bibinfo {author} {\bibfnamefont {N.}~\bibnamefont
  {Korabel}}\ and\ \bibinfo {author} {\bibfnamefont {E.}~\bibnamefont
  {Barkai}},\ }\bibfield  {title} {\enquote {\bibinfo {title} {Pesin-{T}ype
  {I}dentity for {I}ntermittent {D}ynamics with a {Z}ero {L}yapunov
  {E}xponent},}\ }\href {\doibase 10.1103/PhysRevLett.102.050601} {\bibfield
  {journal} {\bibinfo  {journal} {Phys. Rev. Lett.}\ }\textbf {\bibinfo
  {volume} {102}},\ \bibinfo {pages} {050601} (\bibinfo {year}
  {2009})}\BibitemShut {NoStop}%
\bibitem [{\citenamefont {Schulz}\ and\ \citenamefont
  {Barkai}(2015)}]{Schulz:2015}%
  \BibitemOpen
  \bibfield  {author} {\bibinfo {author} {\bibfnamefont {J.~H.~P.}\
  \bibnamefont {Schulz}}\ and\ \bibinfo {author} {\bibfnamefont
  {E.}~\bibnamefont {Barkai}},\ }\bibfield  {title} {\enquote {\bibinfo {title}
  {Fluctuations around equilibrium laws in ergodic continuous-time random
  walks},}\ }\href {\doibase 10.1103/PhysRevE.91.062129} {\bibfield  {journal}
  {\bibinfo  {journal} {Phys. Rev. E}\ }\textbf {\bibinfo {volume} {91}},\
  \bibinfo {pages} {062129} (\bibinfo {year} {2015})}\BibitemShut {NoStop}%
\bibitem [{\citenamefont {Cataliotti}\ \emph {et~al.}(2001)\citenamefont
  {Cataliotti}, \citenamefont {Burger}, \citenamefont {Fort}, \citenamefont
  {Maddaloni}, \citenamefont {Minardi}, \citenamefont {Trombettoni},
  \citenamefont {Smerzi},\ and\ \citenamefont {Inguscio}}]{Cataliotti:2001}%
  \BibitemOpen
  \bibfield  {author} {\bibinfo {author} {\bibfnamefont {F.~S.}\ \bibnamefont
  {Cataliotti}}, \bibinfo {author} {\bibfnamefont {S.}~\bibnamefont {Burger}},
  \bibinfo {author} {\bibfnamefont {C.}~\bibnamefont {Fort}}, \bibinfo {author}
  {\bibfnamefont {P.}~\bibnamefont {Maddaloni}}, \bibinfo {author}
  {\bibfnamefont {F.}~\bibnamefont {Minardi}}, \bibinfo {author} {\bibfnamefont
  {A.}~\bibnamefont {Trombettoni}}, \bibinfo {author} {\bibfnamefont
  {A.}~\bibnamefont {Smerzi}}, \ and\ \bibinfo {author} {\bibfnamefont
  {M.}~\bibnamefont {Inguscio}},\ }\bibfield  {title} {\enquote {\bibinfo
  {title} {{J}osephson {J}unction {A}rrays with {B}ose-{E}instein
  {C}ondensates},}\ }\href {\doibase 10.1126/science.1062612} {\bibfield
  {journal} {\bibinfo  {journal} {Science}\ }\textbf {\bibinfo {volume}
  {293}},\ \bibinfo {pages} {843} (\bibinfo {year} {2001})}\BibitemShut
  {NoStop}%
\bibitem [{\citenamefont {Ryu}\ \emph {et~al.}(2013)\citenamefont {Ryu},
  \citenamefont {Blackburn}, \citenamefont {Blinova},\ and\ \citenamefont
  {Boshier}}]{Ryu:2013}%
  \BibitemOpen
  \bibfield  {author} {\bibinfo {author} {\bibfnamefont {C.}~\bibnamefont
  {Ryu}}, \bibinfo {author} {\bibfnamefont {P.~W.}\ \bibnamefont {Blackburn}},
  \bibinfo {author} {\bibfnamefont {A.~A.}\ \bibnamefont {Blinova}}, \ and\
  \bibinfo {author} {\bibfnamefont {M.~G.}\ \bibnamefont {Boshier}},\
  }\bibfield  {title} {\enquote {\bibinfo {title} {Experimental {R}ealization
  of {J}osephson junctions for an {A}tom {S}{Q}{U}{I}{D}},}\ }\href {\doibase
  10.1103/PhysRevLett.111.205301} {\bibfield  {journal} {\bibinfo  {journal}
  {Phys. Rev. Lett.}\ }\textbf {\bibinfo {volume} {111}},\ \bibinfo {pages}
  {205301} (\bibinfo {year} {2013})}\BibitemShut {NoStop}%
\bibitem [{\citenamefont {Cassidy}\ \emph {et~al.}(2017)\citenamefont
  {Cassidy}, \citenamefont {Bruno}, \citenamefont {Rubbert}, \citenamefont
  {Irfan}, \citenamefont {Kammhuber}, \citenamefont {Schouten}, \citenamefont
  {Akhmerov},\ and\ \citenamefont {Kouwenhoven}}]{Cassidy:2017}%
  \BibitemOpen
  \bibfield  {author} {\bibinfo {author} {\bibfnamefont {M.~C.}\ \bibnamefont
  {Cassidy}}, \bibinfo {author} {\bibfnamefont {A.}~\bibnamefont {Bruno}},
  \bibinfo {author} {\bibfnamefont {S.}~\bibnamefont {Rubbert}}, \bibinfo
  {author} {\bibfnamefont {M.}~\bibnamefont {Irfan}}, \bibinfo {author}
  {\bibfnamefont {J.}~\bibnamefont {Kammhuber}}, \bibinfo {author}
  {\bibfnamefont {R.~N.}\ \bibnamefont {Schouten}}, \bibinfo {author}
  {\bibfnamefont {A.~R.}\ \bibnamefont {Akhmerov}}, \ and\ \bibinfo {author}
  {\bibfnamefont {L.~P.}\ \bibnamefont {Kouwenhoven}},\ }\bibfield  {title}
  {\enquote {\bibinfo {title} {Demonstration of an ac {J}osephson junction
  laser},}\ }\href {\doibase 10.1126/science.aah6640} {\bibfield  {journal}
  {\bibinfo  {journal} {Science}\ }\textbf {\bibinfo {volume} {355}},\ \bibinfo
  {pages} {939} (\bibinfo {year} {2017})}\BibitemShut {NoStop}%
\bibitem [{\citenamefont {Blackburn}\ \emph {et~al.}(2016)\citenamefont
  {Blackburn}, \citenamefont {Cirillo},\ and\ \citenamefont
  {Gr\o{}nbech-Jensen}}]{Blackburn:2016}%
  \BibitemOpen
  \bibfield  {author} {\bibinfo {author} {\bibfnamefont {J.~A.}\ \bibnamefont
  {Blackburn}}, \bibinfo {author} {\bibfnamefont {M.}~\bibnamefont {Cirillo}},
  \ and\ \bibinfo {author} {\bibfnamefont {N.}~\bibnamefont
  {Gr\o{}nbech-Jensen}},\ }\bibfield  {title} {\enquote {\bibinfo {title} {A
  survey of classical and quantum interpretations of experiments on {J}osephson
  junctions at very low temperatures},}\ }\href {\doibase
  https://doi.org/10.1016/j.physrep.2015.10.010} {\bibfield  {journal}
  {\bibinfo  {journal} {Physics Reports}\ }\textbf {\bibinfo {volume} {611}},\
  \bibinfo {pages} {1 } (\bibinfo {year} {2016})}\BibitemShut {NoStop}%
\bibitem [{\citenamefont {Tsygankov}\ and\ \citenamefont
  {Wiesenfeld}(2002)}]{Tsygankov:2002}%
  \BibitemOpen
  \bibfield  {author} {\bibinfo {author} {\bibfnamefont {D.}~\bibnamefont
  {Tsygankov}}\ and\ \bibinfo {author} {\bibfnamefont {K.}~\bibnamefont
  {Wiesenfeld}},\ }\bibfield  {title} {\enquote {\bibinfo {title} {Spontaneous
  synchronization in a {J}osephson transmission line},}\ }\href {\doibase
  10.1103/PhysRevE.66.036215} {\bibfield  {journal} {\bibinfo  {journal} {Phys.
  Rev. E}\ }\textbf {\bibinfo {volume} {66}},\ \bibinfo {pages} {036215}
  (\bibinfo {year} {2002})}\BibitemShut {NoStop}%
\bibitem [{\citenamefont {El-Nashar}\ \emph {et~al.}(2003)\citenamefont
  {El-Nashar}, \citenamefont {Zhang}, \citenamefont {Cerdeira},\ and\
  \citenamefont {Ibiyinka~A.}}]{Nashar:2003}%
  \BibitemOpen
  \bibfield  {author} {\bibinfo {author} {\bibfnamefont {H.~F.}\ \bibnamefont
  {El-Nashar}}, \bibinfo {author} {\bibfnamefont {Y.}~\bibnamefont {Zhang}},
  \bibinfo {author} {\bibfnamefont {H.~A.}\ \bibnamefont {Cerdeira}}, \ and\
  \bibinfo {author} {\bibfnamefont {F.}~\bibnamefont {Ibiyinka~A.}},\
  }\bibfield  {title} {\enquote {\bibinfo {title} {Synchronization in a chain
  of nearest neighbors coupled oscillators with fixed ends},}\ }\href {\doibase
  10.1063/1.1611851} {\bibfield  {journal} {\bibinfo  {journal} {Chaos: An
  Interdisciplinary Journal of Nonlinear Science}\ }\textbf {\bibinfo {volume}
  {13}},\ \bibinfo {pages} {1216} (\bibinfo {year} {2003})}\BibitemShut
  {NoStop}%
\bibitem [{\citenamefont {Binder}\ \emph {et~al.}(2000)\citenamefont {Binder},
  \citenamefont {Abraimov}, \citenamefont {Ustinov}, \citenamefont {Flach},\
  and\ \citenamefont {Zolotaryuk}}]{Binder:2000}%
  \BibitemOpen
  \bibfield  {author} {\bibinfo {author} {\bibfnamefont {P.}~\bibnamefont
  {Binder}}, \bibinfo {author} {\bibfnamefont {D.}~\bibnamefont {Abraimov}},
  \bibinfo {author} {\bibfnamefont {A.~V.}\ \bibnamefont {Ustinov}}, \bibinfo
  {author} {\bibfnamefont {S.}~\bibnamefont {Flach}}, \ and\ \bibinfo {author}
  {\bibfnamefont {Y.}~\bibnamefont {Zolotaryuk}},\ }\bibfield  {title}
  {\enquote {\bibinfo {title} {Observation of breathers in {J}osephson
  ladders},}\ }\href {\doibase 10.1103/PhysRevLett.84.745} {\bibfield
  {journal} {\bibinfo  {journal} {Phys. Rev. Lett.}\ }\textbf {\bibinfo
  {volume} {84}},\ \bibinfo {pages} {745} (\bibinfo {year} {2000})}\BibitemShut
  {NoStop}%
\bibitem [{\citenamefont {Miroshnichenko}\ \emph {et~al.}(2001)\citenamefont
  {Miroshnichenko}, \citenamefont {Flach}, \citenamefont {Fistul},
  \citenamefont {Zolotaryuk},\ and\ \citenamefont
  {Page}}]{Miroshnichenko:2001}%
  \BibitemOpen
  \bibfield  {author} {\bibinfo {author} {\bibfnamefont {A.~E.}\ \bibnamefont
  {Miroshnichenko}}, \bibinfo {author} {\bibfnamefont {S.}~\bibnamefont
  {Flach}}, \bibinfo {author} {\bibfnamefont {M.~V.}\ \bibnamefont {Fistul}},
  \bibinfo {author} {\bibfnamefont {Y.}~\bibnamefont {Zolotaryuk}}, \ and\
  \bibinfo {author} {\bibfnamefont {J.~B.}\ \bibnamefont {Page}},\ }\bibfield
  {title} {\enquote {\bibinfo {title} {Breathers in {J}osephson junction
  ladders: {R}esonances and electromagnetic wave spectroscopy},}\ }\href
  {\doibase 10.1103/PhysRevE.64.066601} {\bibfield  {journal} {\bibinfo
  {journal} {Phys. Rev. E}\ }\textbf {\bibinfo {volume} {64}},\ \bibinfo
  {pages} {066601} (\bibinfo {year} {2001})}\BibitemShut {NoStop}%
\bibitem [{\citenamefont {Fistul}\ \emph {et~al.}(2002)\citenamefont {Fistul},
  \citenamefont {Miroshnichenko}, \citenamefont {Flach}, \citenamefont
  {Schuster},\ and\ \citenamefont {Ustinov}}]{Fistul:2002}%
  \BibitemOpen
  \bibfield  {author} {\bibinfo {author} {\bibfnamefont {M.~V.}\ \bibnamefont
  {Fistul}}, \bibinfo {author} {\bibfnamefont {A.~E.}\ \bibnamefont
  {Miroshnichenko}}, \bibinfo {author} {\bibfnamefont {S.}~\bibnamefont
  {Flach}}, \bibinfo {author} {\bibfnamefont {M.}~\bibnamefont {Schuster}}, \
  and\ \bibinfo {author} {\bibfnamefont {A.~V.}\ \bibnamefont {Ustinov}},\
  }\bibfield  {title} {\enquote {\bibinfo {title} {Incommensurate dynamics of
  resonant breathers in {J}osephson junction ladders},}\ }\href {\doibase
  10.1103/PhysRevB.65.174524} {\bibfield  {journal} {\bibinfo  {journal} {Phys.
  Rev. B}\ }\textbf {\bibinfo {volume} {65}},\ \bibinfo {pages} {174524}
  (\bibinfo {year} {2002})}\BibitemShut {NoStop}%
\bibitem [{\citenamefont {Miroshnichenko}\ \emph {et~al.}(2005)\citenamefont
  {Miroshnichenko}, \citenamefont {Schuster}, \citenamefont {Flach},
  \citenamefont {Fistul},\ and\ \citenamefont {Ustinov}}]{Miroshnichenko:2005}%
  \BibitemOpen
  \bibfield  {author} {\bibinfo {author} {\bibfnamefont {A.~E.}\ \bibnamefont
  {Miroshnichenko}}, \bibinfo {author} {\bibfnamefont {M.}~\bibnamefont
  {Schuster}}, \bibinfo {author} {\bibfnamefont {S.}~\bibnamefont {Flach}},
  \bibinfo {author} {\bibfnamefont {M.~V.}\ \bibnamefont {Fistul}}, \ and\
  \bibinfo {author} {\bibfnamefont {A.~V.}\ \bibnamefont {Ustinov}},\
  }\bibfield  {title} {\enquote {\bibinfo {title} {Resonant plasmon scattering
  by discrete breathers in {J}osephson junction ladders},}\ }\href {\doibase
  10.1103/PhysRevB.71.174306} {\bibfield  {journal} {\bibinfo  {journal} {Phys.
  Rev. B}\ }\textbf {\bibinfo {volume} {71}},\ \bibinfo {pages} {174306}
  (\bibinfo {year} {2005})}\BibitemShut {NoStop}%
\bibitem [{\citenamefont {Shulga}\ \emph {et~al.}(2018)\citenamefont {Shulga},
  \citenamefont {Il'ichev}, \citenamefont {Fistul}, \citenamefont {Besedin},
  \citenamefont {Butz}, \citenamefont {Astafiev}, \citenamefont {H\"ubner},\
  and\ \citenamefont {Ustinov}}]{Shulga:2018}%
  \BibitemOpen
  \bibfield  {author} {\bibinfo {author} {\bibfnamefont {K.~V.}\ \bibnamefont
  {Shulga}}, \bibinfo {author} {\bibfnamefont {E.}~\bibnamefont {Il'ichev}},
  \bibinfo {author} {\bibfnamefont {M.~V.}\ \bibnamefont {Fistul}}, \bibinfo
  {author} {\bibfnamefont {I.~S.}\ \bibnamefont {Besedin}}, \bibinfo {author}
  {\bibfnamefont {S.}~\bibnamefont {Butz}}, \bibinfo {author} {\bibfnamefont
  {O.~V.}\ \bibnamefont {Astafiev}}, \bibinfo {author} {\bibfnamefont
  {U.}~\bibnamefont {H\"ubner}}, \ and\ \bibinfo {author} {\bibfnamefont
  {A.~V.}\ \bibnamefont {Ustinov}},\ }\bibfield  {title} {\enquote {\bibinfo
  {title} {Magnetically induced transparency of a quantum metamaterial composed
  of twin flux qubits},}\ }\href {https://doi.org/10.1038/s41467-017-02608-8}
  {\bibfield  {journal} {\bibinfo  {journal} {Nature Communications}\ }\textbf
  {\bibinfo {volume} {9}},\ \bibinfo {pages} {150} (\bibinfo {year}
  {2018})}\BibitemShut {NoStop}%
\bibitem [{\citenamefont {Martinis}(2004)}]{Martinis:2004}%
  \BibitemOpen
  \bibfield  {author} {\bibinfo {author} {\bibfnamefont {J.~M.}\ \bibnamefont
  {Martinis}},\ }\bibfield  {title} {\enquote {\bibinfo {title} {Course 13 -
  {S}uperconducting {Q}ubits and the {P}hysics of {J}osephson {J}unctions},}\
  }in\ \href {\doibase https://doi.org/10.1016/S0924-8099(03)80037-9} {\emph
  {\bibinfo {booktitle} {Quantum Entanglement and Information Processing}}},\
  \bibinfo {series} {Les Houches}, Vol.~\bibinfo {volume} {79},\ \bibinfo
  {editor} {edited by\ \bibinfo {editor} {\bibfnamefont {D.}~\bibnamefont
  {Est\'eve}}, \bibinfo {editor} {\bibfnamefont {J.-M.}\ \bibnamefont
  {Raimond}}, \ and\ \bibinfo {editor} {\bibfnamefont {J.}~\bibnamefont
  {Dalibard}}}\ (\bibinfo  {publisher} {Elsevier},\ \bibinfo {year} {2004})\
  pp.\ \bibinfo {pages} {487 -- 520}\BibitemShut {NoStop}%
\bibitem [{\citenamefont {Gendelman}\ and\ \citenamefont
  {Savin}(2000)}]{Gendelman:2000}%
  \BibitemOpen
  \bibfield  {author} {\bibinfo {author} {\bibfnamefont {O.~V.}\ \bibnamefont
  {Gendelman}}\ and\ \bibinfo {author} {\bibfnamefont {A.~V.}\ \bibnamefont
  {Savin}},\ }\bibfield  {title} {\enquote {\bibinfo {title} {Normal {H}eat
  {C}onductivity of the {O}ne-{D}imensional {L}attice with {P}eriodic
  {P}otential of {N}earest-{N}eighbor {I}nteraction},}\ }\href {\doibase
  10.1103/PhysRevLett.84.2381} {\bibfield  {journal} {\bibinfo  {journal}
  {Phys. Rev. Lett.}\ }\textbf {\bibinfo {volume} {84}},\ \bibinfo {pages}
  {2381} (\bibinfo {year} {2000})}\BibitemShut {NoStop}%
\bibitem [{\citenamefont {Giardin\`a}\ \emph {et~al.}(2000)\citenamefont
  {Giardin\`a}, \citenamefont {Livi}, \citenamefont {Politi},\ and\
  \citenamefont {Vassalli}}]{Giardina:2000}%
  \BibitemOpen
  \bibfield  {author} {\bibinfo {author} {\bibfnamefont {C.}~\bibnamefont
  {Giardin\`a}}, \bibinfo {author} {\bibfnamefont {R.}~\bibnamefont {Livi}},
  \bibinfo {author} {\bibfnamefont {A.}~\bibnamefont {Politi}}, \ and\ \bibinfo
  {author} {\bibfnamefont {M.}~\bibnamefont {Vassalli}},\ }\bibfield  {title}
  {\enquote {\bibinfo {title} {Finite {T}hermal {C}onductivity in 1{D}
  {L}attices},}\ }\href {\doibase 10.1103/PhysRevLett.84.2144} {\bibfield
  {journal} {\bibinfo  {journal} {Phys. Rev. Lett.}\ }\textbf {\bibinfo
  {volume} {84}},\ \bibinfo {pages} {2144} (\bibinfo {year}
  {2000})}\BibitemShut {NoStop}%
\bibitem [{\citenamefont {Pino}\ \emph {et~al.}(2016)\citenamefont {Pino},
  \citenamefont {Ioffe},\ and\ \citenamefont {Altshuler}}]{Pino:2016}%
  \BibitemOpen
  \bibfield  {author} {\bibinfo {author} {\bibfnamefont {M.}~\bibnamefont
  {Pino}}, \bibinfo {author} {\bibfnamefont {L.~B.}\ \bibnamefont {Ioffe}}, \
  and\ \bibinfo {author} {\bibfnamefont {B.~L.}\ \bibnamefont {Altshuler}},\
  }\bibfield  {title} {\enquote {\bibinfo {title} {Nonergodic metallic and
  insulating phases of {J}osephson junction chains},}\ }\href {\doibase
  10.1073/pnas.1520033113} {\bibfield  {journal} {\bibinfo  {journal} {PNAS}\
  }\textbf {\bibinfo {volume} {113}},\ \bibinfo {pages} {536} (\bibinfo {year}
  {2016})}\BibitemShut {NoStop}%
\bibitem [{\citenamefont {Basko}\ \emph {et~al.}(2006)\citenamefont {Basko},
  \citenamefont {Aleiner},\ and\ \citenamefont {Altshuler}}]{Basko2006metal}%
  \BibitemOpen
  \bibfield  {author} {\bibinfo {author} {\bibfnamefont {D.}~\bibnamefont
  {Basko}}, \bibinfo {author} {\bibfnamefont {I.}~\bibnamefont {Aleiner}}, \
  and\ \bibinfo {author} {\bibfnamefont {B.}~\bibnamefont {Altshuler}},\
  }\bibfield  {title} {\enquote {\bibinfo {title} {Metal insulator transition
  in a weakly interacting many-electron system with localized single-particle
  states},}\ }\href {\doibase https://doi.org/10.1016/j.aop.2005.11.014}
  {\bibfield  {journal} {\bibinfo  {journal} {Annals of Physics}\ }\textbf
  {\bibinfo {volume} {321}},\ \bibinfo {pages} {1126 } (\bibinfo {year}
  {2006})}\BibitemShut {NoStop}%
\bibitem [{\citenamefont {Escande}\ \emph {et~al.}(1994)\citenamefont
  {Escande}, \citenamefont {Kantz}, \citenamefont {Livi},\ and\ \citenamefont
  {Ruffo}}]{Escande:1994}%
  \BibitemOpen
  \bibfield  {author} {\bibinfo {author} {\bibfnamefont {D.}~\bibnamefont
  {Escande}}, \bibinfo {author} {\bibfnamefont {H.}~\bibnamefont {Kantz}},
  \bibinfo {author} {\bibfnamefont {R.}~\bibnamefont {Livi}}, \ and\ \bibinfo
  {author} {\bibfnamefont {S.}~\bibnamefont {Ruffo}},\ }\bibfield  {title}
  {\enquote {\bibinfo {title} {Self-consistent check of the validity of {G}ibbs
  calculus using dynamical variables},}\ }\href {\doibase 10.1007/BF02188677}
  {\bibfield  {journal} {\bibinfo  {journal} {Journal of Statistical Physics}\
  }\textbf {\bibinfo {volume} {76}},\ \bibinfo {pages} {605} (\bibinfo {year}
  {1994})}\BibitemShut {NoStop}%
\bibitem [{\citenamefont {De~Roeck}\ and\ \citenamefont
  {Huveneers}(2014)}]{Roeck:2014}%
  \BibitemOpen
  \bibfield  {author} {\bibinfo {author} {\bibfnamefont {W.}~\bibnamefont
  {De~Roeck}}\ and\ \bibinfo {author} {\bibfnamefont {F.}~\bibnamefont
  {Huveneers}},\ }\bibfield  {title} {\enquote {\bibinfo {title} {Asymptotic
  {L}ocalization of {E}nergy in {N}ondisordered {O}scillator {C}hains},}\
  }\href {\doibase 10.1002/cpa.21550} {\bibfield  {journal} {\bibinfo
  {journal} {Communications on Pure and Applied Mathematics}\ }\textbf
  {\bibinfo {volume} {68}},\ \bibinfo {pages} {1532} (\bibinfo {year}
  {2014})}\BibitemShut {NoStop}%
\bibitem [{\citenamefont {Mithun}\ \emph {et~al.}(2018)\citenamefont {Mithun},
  \citenamefont {Kati}, \citenamefont {Danieli},\ and\ \citenamefont
  {Flach}}]{Mithun:2018}%
  \BibitemOpen
  \bibfield  {author} {\bibinfo {author} {\bibfnamefont {T.}~\bibnamefont
  {Mithun}}, \bibinfo {author} {\bibfnamefont {Y.}~\bibnamefont {Kati}},
  \bibinfo {author} {\bibfnamefont {C.}~\bibnamefont {Danieli}}, \ and\
  \bibinfo {author} {\bibfnamefont {S.}~\bibnamefont {Flach}},\ }\bibfield
  {title} {\enquote {\bibinfo {title} {Weakly {N}onergodic {D}ynamics in the
  {G}ross-{P}itaevskii {L}attice},}\ }\href {\doibase
  10.1103/PhysRevLett.120.184101} {\bibfield  {journal} {\bibinfo  {journal}
  {Phys. Rev. Lett.}\ }\textbf {\bibinfo {volume} {120}},\ \bibinfo {pages}
  {184101} (\bibinfo {year} {2018})}\BibitemShut {NoStop}%
\bibitem [{\citenamefont {Danieli}\ \emph {et~al.}(2017)\citenamefont
  {Danieli}, \citenamefont {Campbell},\ and\ \citenamefont
  {Flach}}]{Danieli:2017}%
  \BibitemOpen
  \bibfield  {author} {\bibinfo {author} {\bibfnamefont {C.}~\bibnamefont
  {Danieli}}, \bibinfo {author} {\bibfnamefont {D.~K.}\ \bibnamefont
  {Campbell}}, \ and\ \bibinfo {author} {\bibfnamefont {S.}~\bibnamefont
  {Flach}},\ }\bibfield  {title} {\enquote {\bibinfo {title} {Intermittent
  many-body dynamics at equilibrium},}\ }\href {\doibase
  10.1103/PhysRevE.95.060202} {\bibfield  {journal} {\bibinfo  {journal} {Phys.
  Rev. E}\ }\textbf {\bibinfo {volume} {95}},\ \bibinfo {pages} {060202}
  (\bibinfo {year} {2017})}\BibitemShut {NoStop}%
\bibitem [{\citenamefont {Casetti}\ \emph {et~al.}(1996)\citenamefont
  {Casetti}, \citenamefont {Clementi},\ and\ \citenamefont
  {Pettini}}]{Casetti:1996}%
  \BibitemOpen
  \bibfield  {author} {\bibinfo {author} {\bibfnamefont {L.}~\bibnamefont
  {Casetti}}, \bibinfo {author} {\bibfnamefont {C.}~\bibnamefont {Clementi}}, \
  and\ \bibinfo {author} {\bibfnamefont {M.}~\bibnamefont {Pettini}},\
  }\bibfield  {title} {\enquote {\bibinfo {title} {Riemannian theory of
  {H}amiltonian chaos and {L}yapunov exponents},}\ }\href {\doibase
  10.1103/PhysRevE.54.5969} {\bibfield  {journal} {\bibinfo  {journal} {Phys.
  Rev. E}\ }\textbf {\bibinfo {volume} {54}},\ \bibinfo {pages} {5969}
  (\bibinfo {year} {1996})}\BibitemShut {NoStop}%
\bibitem [{\citenamefont {Casetti}\ \emph {et~al.}(2000)\citenamefont
  {Casetti}, \citenamefont {Pettini},\ and\ \citenamefont
  {Cohen}}]{Casetti:2000}%
  \BibitemOpen
  \bibfield  {author} {\bibinfo {author} {\bibfnamefont {L.}~\bibnamefont
  {Casetti}}, \bibinfo {author} {\bibfnamefont {M.}~\bibnamefont {Pettini}}, \
  and\ \bibinfo {author} {\bibfnamefont {E.}~\bibnamefont {Cohen}},\ }\bibfield
   {title} {\enquote {\bibinfo {title} {Geometric approach to {H}amiltonian
  dynamics and statistical mechanics},}\ }\href {\doibase
  https://doi.org/10.1016/S0370-1573(00)00069-7} {\bibfield  {journal}
  {\bibinfo  {journal} {Physics Reports}\ }\textbf {\bibinfo {volume} {337}},\
  \bibinfo {pages} {237 } (\bibinfo {year} {2000})}\BibitemShut {NoStop}%
\bibitem [{\citenamefont {Livi}\ \emph {et~al.}(1987)\citenamefont {Livi},
  \citenamefont {Pettini}, \citenamefont {Ruffo},\ and\ \citenamefont
  {Vulpiani}}]{Livi:1987}%
  \BibitemOpen
  \bibfield  {author} {\bibinfo {author} {\bibfnamefont {R.}~\bibnamefont
  {Livi}}, \bibinfo {author} {\bibfnamefont {M.}~\bibnamefont {Pettini}},
  \bibinfo {author} {\bibfnamefont {S.}~\bibnamefont {Ruffo}}, \ and\ \bibinfo
  {author} {\bibfnamefont {A.}~\bibnamefont {Vulpiani}},\ }\bibfield  {title}
  {\enquote {\bibinfo {title} {Chaotic behavior in nonlinear {H}amiltonian
  systems and equilibrium statistical mechanics},}\ }\href@noop {} {\bibfield
  {journal} {\bibinfo  {journal} {Journal of Statistical Physics}\ }\textbf
  {\bibinfo {volume} {48}},\ \bibinfo {pages} {539} (\bibinfo {year}
  {1987})}\BibitemShut {NoStop}%
\bibitem [{Sup()}]{Supple}%
  \BibitemOpen
  \href@noop {} {\bibinfo  {journal} {See Supplemental Material at [URL will be
  inserted by publisher] for additional information, which include
  Refs.\cite{Skokos:2009,richard1972feynman,Chakravarty1985photoinduced,kac1963van}}\
  }\BibitemShut {NoStop}%
\bibitem [{\citenamefont {Chirikov}(1979)}]{Chirikov:1979}%
  \BibitemOpen
\bibfield  {journal} {  }\bibfield  {author} {\bibinfo {author} {\bibfnamefont
  {B.~V.}\ \bibnamefont {Chirikov}},\ }\bibfield  {title} {\enquote {\bibinfo
  {title} {A universal instability of many-dimensional oscillator systems},}\
  }\href@noop {} {\bibfield  {journal} {\bibinfo  {journal} {Physics Reports}\
  }\textbf {\bibinfo {volume} {52}},\ \bibinfo {pages} {263} (\bibinfo {year}
  {1979})}\BibitemShut {NoStop}%
\bibitem [{\citenamefont {Kolmogorov}(1954)}]{Kolmogorov1954conservation}%
  \BibitemOpen
  \bibfield  {author} {\bibinfo {author} {\bibfnamefont {A.~N.}\ \bibnamefont
  {Kolmogorov}},\ }\bibfield  {title} {\enquote {\bibinfo {title} {{On
  conservation of conditionally periodic motions for a small change in
  Hamilton's function}},}\ }\href {http://cds.cern.ch/record/430016} {\bibfield
   {journal} {\bibinfo  {journal} {Dokl. Akad. Nauk SSSR}\ }\textbf {\bibinfo
  {volume} {98}},\ \bibinfo {pages} {527} (\bibinfo {year} {1954})}\BibitemShut
  {NoStop}%
\bibitem [{\citenamefont {Arnold}(1963)}]{arnold1963aproof}%
  \BibitemOpen
  \bibfield  {author} {\bibinfo {author} {\bibfnamefont {V.}~\bibnamefont
  {Arnold}},\ }\bibfield  {title} {\enquote {\bibinfo {title} {A proof of a
  theorem by a.n. kolmogorov on the invariance of quasi-periodic motions under
  small perturbations of the hamiltonian},}\ }\href@noop {} {\bibfield
  {journal} {\bibinfo  {journal} {Russ. Math. Surv.}\ }\textbf {\bibinfo
  {volume} {18}},\ \bibinfo {pages} {9} (\bibinfo {year} {1963})}\BibitemShut
  {NoStop}%
\bibitem [{\citenamefont {Moser}(1962)}]{Moser1962invariant}%
  \BibitemOpen
  \bibfield  {author} {\bibinfo {author} {\bibfnamefont {J.}~\bibnamefont
  {Moser}},\ }\bibfield  {title} {\enquote {\bibinfo {title} {{On invariant
  curves of area-preserving mappings of an annulus}},}\ }\href
  {http://cds.cern.ch/record/430015} {\bibfield  {journal} {\bibinfo  {journal}
  {Nachr. Akad. Wiss. Gottingen, II}\ ,\ \bibinfo {pages} {1}} (\bibinfo {year}
  {1962})}\BibitemShut {NoStop}%
\bibitem [{\citenamefont {Chierchia}\ and\ \citenamefont
  {Gallavotti}(1982)}]{Chierchia1982smooth}%
  \BibitemOpen
  \bibfield  {author} {\bibinfo {author} {\bibfnamefont {L.}~\bibnamefont
  {Chierchia}}\ and\ \bibinfo {author} {\bibfnamefont {G.}~\bibnamefont
  {Gallavotti}},\ }\bibfield  {title} {\enquote {\bibinfo {title} {Smooth prime
  integrals for quasi-integrable hamiltonian systems},}\ }\href {\doibase
  10.1007/BF02721167} {\bibfield  {journal} {\bibinfo  {journal} {Il Nuovo
  Cimento B}\ }\textbf {\bibinfo {volume} {67}},\ \bibinfo {pages} {277}
  (\bibinfo {year} {1982})}\BibitemShut {NoStop}%
\bibitem [{\citenamefont {Benettin}\ \emph {et~al.}(1984)\citenamefont
  {Benettin}, \citenamefont {Galgani}, \citenamefont {Giorgilli},\ and\
  \citenamefont {Strelcyn}}]{Benettin1984proof}%
  \BibitemOpen
  \bibfield  {author} {\bibinfo {author} {\bibfnamefont {G.}~\bibnamefont
  {Benettin}}, \bibinfo {author} {\bibfnamefont {L.}~\bibnamefont {Galgani}},
  \bibinfo {author} {\bibfnamefont {A.}~\bibnamefont {Giorgilli}}, \ and\
  \bibinfo {author} {\bibfnamefont {J.~M.}\ \bibnamefont {Strelcyn}},\
  }\bibfield  {title} {\enquote {\bibinfo {title} {A proof of kolmogorov's
  theorem on invariant tori using canonical transformations defined by the lie
  method},}\ }\href {\doibase 10.1007/BF02748972} {\bibfield  {journal}
  {\bibinfo  {journal} {Il Nuovo Cimento B}\ }\textbf {\bibinfo {volume}
  {79}},\ \bibinfo {pages} {201} (\bibinfo {year} {1984})}\BibitemShut
  {NoStop}%
\bibitem [{\citenamefont {Wayne}(1984{\natexlab{a}})}]{Wayne1984KAM1}%
  \BibitemOpen
  \bibfield  {author} {\bibinfo {author} {\bibfnamefont {C.~E.}\ \bibnamefont
  {Wayne}},\ }\bibfield  {title} {\enquote {\bibinfo {title} {The kam theory of
  systems with short range interactions, 1},}\ }\href {\doibase
  10.1007/BF01214577} {\bibfield  {journal} {\bibinfo  {journal} {Comm. Math.
  Phys.}\ }\textbf {\bibinfo {volume} {96}},\ \bibinfo {pages} {311} (\bibinfo
  {year} {1984}{\natexlab{a}})}\BibitemShut {NoStop}%
\bibitem [{\citenamefont {Wayne}(1984{\natexlab{b}})}]{Wayne1984KAM2}%
  \BibitemOpen
  \bibfield  {author} {\bibinfo {author} {\bibfnamefont {C.~E.}\ \bibnamefont
  {Wayne}},\ }\bibfield  {title} {\enquote {\bibinfo {title} {The kam theory of
  systems with short range interactions, 2},}\ }\href {\doibase
  10.1007/BF01214578} {\bibfield  {journal} {\bibinfo  {journal} {Comm. Math.
  Phys.}\ }\textbf {\bibinfo {volume} {96}},\ \bibinfo {pages} {331} (\bibinfo
  {year} {1984}{\natexlab{b}})}\BibitemShut {NoStop}%
\bibitem [{\citenamefont {Skokos}\ \emph {et~al.}(2009)\citenamefont {Skokos},
  \citenamefont {Krimer}, \citenamefont {Komineas},\ and\ \citenamefont
  {Flach}}]{Skokos:2009}%
  \BibitemOpen
  \bibfield  {author} {\bibinfo {author} {\bibfnamefont {C.}~\bibnamefont
  {Skokos}}, \bibinfo {author} {\bibfnamefont {D.~O.}\ \bibnamefont {Krimer}},
  \bibinfo {author} {\bibfnamefont {S.}~\bibnamefont {Komineas}}, \ and\
  \bibinfo {author} {\bibfnamefont {S.}~\bibnamefont {Flach}},\ }\bibfield
  {title} {\enquote {\bibinfo {title} {Delocalization of wave packets in
  disordered nonlinear chains},}\ }\href {\doibase 10.1103/PhysRevE.79.056211}
  {\bibfield  {journal} {\bibinfo  {journal} {Phys. Rev. E}\ }\textbf {\bibinfo
  {volume} {79}},\ \bibinfo {pages} {056211} (\bibinfo {year}
  {2009})}\BibitemShut {NoStop}%
\bibitem [{\citenamefont {Richard}(1972)}]{richard1972feynman}%
  \BibitemOpen
  \bibfield  {author} {\bibinfo {author} {\bibfnamefont {P.}~\bibnamefont
  {Richard}},\ }\bibfield  {title} {\enquote {\bibinfo {title} {Feynman.
  statistical {M}echanics, a set of lectures},}\ }\href@noop {} {\bibfield
  {journal} {\bibinfo  {journal} {Frontiers in Physics. Perseus Books}\ }
  (\bibinfo {year} {1972})}\BibitemShut {NoStop}%
\bibitem [{\citenamefont {Chakravarty}\ and\ \citenamefont
  {Kivelson}(1985)}]{Chakravarty1985photoinduced}%
  \BibitemOpen
  \bibfield  {author} {\bibinfo {author} {\bibfnamefont {S.}~\bibnamefont
  {Chakravarty}}\ and\ \bibinfo {author} {\bibfnamefont {S.}~\bibnamefont
  {Kivelson}},\ }\bibfield  {title} {\enquote {\bibinfo {title} {Photoinduced
  macroscopic quantum tunneling},}\ }\href {\doibase 10.1103/PhysRevB.32.76}
  {\bibfield  {journal} {\bibinfo  {journal} {Phys. Rev. B}\ }\textbf {\bibinfo
  {volume} {32}},\ \bibinfo {pages} {76} (\bibinfo {year} {1985})}\BibitemShut
  {NoStop}%
\bibitem [{\citenamefont {Kac}\ \emph {et~al.}(1963)\citenamefont {Kac},
  \citenamefont {Uhlenbeck},\ and\ \citenamefont {Hemmer}}]{kac1963van}%
  \BibitemOpen
  \bibfield  {author} {\bibinfo {author} {\bibfnamefont {M.}~\bibnamefont
  {Kac}}, \bibinfo {author} {\bibfnamefont {G.}~\bibnamefont {Uhlenbeck}}, \
  and\ \bibinfo {author} {\bibfnamefont {P.}~\bibnamefont {Hemmer}},\
  }\bibfield  {title} {\enquote {\bibinfo {title} {On the van der {W}aals
  {T}heory of the {V}apor-{L}iquid {E}quilibrium. i. {D}iscussion of a
  {O}ne-{D}imensional {M}odel},}\ }\href@noop {} {\bibfield  {journal}
  {\bibinfo  {journal} {Journal of Mathematical Physics}\ }\textbf {\bibinfo
  {volume} {4}},\ \bibinfo {pages} {216} (\bibinfo {year} {1963})}\BibitemShut
  {NoStop}%
\end{thebibliography}%

\clearpage

\section*{Supplemental Material}
 \section{Statistical Analysis} \label{sup1}
 The energy density $h$  is calculated with the microcanonical partition function
    \begin{equation}
    Z =\int_{-\infty}^{\infty} \int_{-\pi}^{\pi}\prod_{n} dp_n dq_n e^{-\beta H}
    \label{eq:rotor1a}
\end{equation}
as
    \begin{equation}
h=-\frac{1}{N}\frac{\partial \ln(Z)}{\partial \beta}=\frac{1}{2\beta}+ E_J\big(1-  \frac{I_1(\beta E_J)}{I_0(\beta E_J)}\big),
    \label{eq:rotor1b}
\end{equation}
with average potential energy density 
    \begin{equation}
            \begin{split}
u&=E_J\big(1-\frac{I_1(\beta E_J)}{I_0(\beta E_J)}\big)
    \label{eq:rotor1c}
     \end{split}
\end{equation}
and 
average kinetic energy density 
    \begin{equation}
            \begin{split}
k&=\frac{1}{2\beta}.
    \label{eq:rotor1d}
     \end{split}
\end{equation}
In terms of $k$ we rewrite Eq.~\ref{eq:rotor1b} as
\begin{equation}
h = k +E_J\left( 1 - \frac{I_1(E_J/2k)}{I_0(E_J/2k)} \right).
    \label{eq:rotor1e}
\end{equation}

  \section{Integration}\label{ap1}
We split Eq.~1 in the main text as 
  \begin{equation}
\small  A = \sum_{n=1}^{N}\frac{p_n^2}{2}\ ,\qquad B = E_J \sum_{n=1}^{N} (1-\cos(q_{n+1}-q_n))  \;.
\label{eq:a1}
\end{equation}
As discussed in \cite{Skokos:2009}, this separation leads to a symplectic integration scheme called $\text{SBAB}_2$, where
\begin{equation}
\begin{split}
e^{\Delta t \mathcal{H}}&=e^{\Delta t (A+B)}\approx e^{d_1\Delta t L_B}e^{c_2\Delta t L_A}e^{d_2\Delta t L_B}\\ \times & e^{c_2\Delta t L_A} e^{d_1\Delta t L_B}  \\
\end{split}
\label{eq:a2}
\end{equation}
where $d_1=\frac{1}{6}$, $d_2=\frac{2}{3}$, $c_2=\frac{1}{2}$. 
The operators $e^{\Delta t L_A}$ and $e^{\Delta t L_B}$ which propagate the set of initial conditions ($q_n,p_n$) from Eq.~(\ref{eq:a1}) at the time $t$ to the final values ($q_n^{'},p_n^{'}$) at the time $t+\Delta t$ are 
  \begin{equation}
   \begin{split}
    e^{\Delta t L_A}&:\left\{
                \begin{array}{ll}
                  q_n^{'} =q_n+ p_n \Delta t\\
                   p_n^{'} =p_n
                \end{array}
              \right. \\
                  e^{\Delta t L_B}&:\left\{
          \begin{array}{ll}
              q_n^{'} =q_n\\
                p_n^{'} =p_n +  E_J \big[  \sin(q_{n+1}-q_n)+\sin(q_{n-1} - q_n)  \big] \Delta t
          \end{array}
          \right.
              \label{eq:a4}
     \end{split}
  \end{equation}
We then introduce a corrector $C=\{ \{A,B\} , B\}$. Following \cite{Skokos:2009}, this term applies 
\begin{equation}
\begin{split}
\text{SBAB}_2 C  =  e^{-\frac{g}{2}\Delta t^3 L_C}  \text{SBAB}_2    e^{-\frac{g}{2}\Delta t^3 L_C}
\end{split}
\label{eq:a5}
\end{equation}
for $g = 1/72$. The corrector term is 
\begin{equation}
\begin{split}
C &= - \sum_{n=1}^{N}    \frac{\partial \{ A,B\}}{\partial p_n}\frac{\partial B}{\partial q_n}=    \sum_{n=1}^{N} \bigg( \frac{\partial B}{\partial q_n}  \bigg)^2 \\
\quad &=  E_J^2 \sum_{n=1}^{N} \big[  \sin(q_{n+1}-q_n)+\sin(q_{n-1} - q_n)  \big]^2.
\label{eq:a7}
\end{split}
\end{equation}
The corrector operator $C$ yields to the following resolvent operator 
      \begin{widetext}
  \begin{equation*}
                  e^{t L_C}:\left\{
          \begin{array}{ll}
              q_n^{'} =q_n\\
                p_n^{'} =p_n +  E_J^2 \Big\{ 2 \big[  \sin(q_{n+1}-q_n)+\sin(q_{n-1} - q_n)  \big] \cdot \big[ \cos(q_{n+1}-q_n) + \cos(q_{n-1} - q_n)  \big] \\
\qquad -2 \big[  \sin(q_{n+2}-q_{n+1})+\sin(q_{n} - q_{n+1})  \big] \cdot  \cos(q_{n} - q_{n+1})   \\
\qquad - 2 \big[  \sin(q_{n}-q_{n-1})+\sin(q_{n-2} - q_{n-1})  \big] \cdot   \cos(q_{n} - q_{n-1})  \Big\}   \Delta t
          \end{array}
          \right.
              \label{eq:a10}
  \end{equation*}
 \end{widetext}
    \section{Calculation of maximal LCE : Tangent map method and Variational Equations} \label{ap2}
    If the autonomous Hamiltonian has the form \cite{Skokos:2009}
    \begin{equation}
 H(\vec{q},\vec{p}) =\sum_{n=1}^{N}\bigg[\frac{1}{2}\vec{p}_n^2+V(\vec{q}) \bigg],
\label{eq:b1}
\end{equation}
the equations of motion are 
 \begin{equation}
\begin{bmatrix}
    \dot{\vec{q}}  \\
    \dot{\vec{p}} \\
\end{bmatrix}
=
\begin{bmatrix}
   \vec{p} \\
   -\frac{\partial V(\vec{q})}{\partial \vec{q}}
\end{bmatrix}
\end{equation}
The corresponding variational Hamiltonian and equations of motion are
    \begin{equation}
  H_V(\vec{\delta q},\vec{\delta p}) =\sum_{n=1}^{N}\bigg[\frac{1}{2}\delta \vec{p}_n^2+\frac{1}{2} \sum_{m=1}^{N} D_V^2(\vec{q})_{nm}\delta \vec{q}_n \delta \vec{q}_m \bigg],
\label{eq:b2}
\end{equation}
 \begin{equation}
\text{and}~~~\begin{bmatrix}
    \delta \dot{\vec{q}}  \\
    \delta \dot{\vec{p}} \\
\end{bmatrix}
=
\begin{bmatrix}
   \delta \vec{p} \\
   -D_V^2(\vec{q})\delta \vec{q}
\end{bmatrix},
\end{equation}
respectively. Here,
 \begin{equation}
D_V^2(\vec{q(t)})_{nm}=\frac{\partial^2 V(\vec{q})}{\partial \vec{q}_n \vec{q}_m}|_{\vec{q}(t)}
\end{equation}
For Eq.(1) in the main text, the variational equations of motion are 
  \begin{widetext}
 \begin{equation}
\begin{bmatrix}
    \delta \dot{q_n}  \\
    \delta \dot{p_n} \\
\end{bmatrix}
=
\begin{bmatrix}
   \delta p_n \\
   -E_J\big[-\cos(q_n-q_{n-1}) \delta q_{n-1}+(\cos(q_{n+1}-q_n)+\cos(q_n-q_{n-1}))\delta q_n-\cos(q_{n+1}-q_n) \delta q_{n+1} \big]
\end{bmatrix},
\end{equation}
  \end{widetext}
The corresponding operators are 
 \begin{equation}
 \begin{split}
    e^{\Delta t L_{AV}}&:\left\{
                \begin{array}{ll}
                  \vec{\delta} q^{'} =\vec{\delta} q+ \vec{\delta} p \Delta t\\
                   \vec{\delta} p^{'} =\vec{\delta} p
                \end{array}
              \right. \\
              \qquad \qquad 
                  e^{\Delta t L_{BV}}&:\left\{
          \begin{array}{ll}
             \vec{\delta} q^{'} =\vec{\delta} q\\
              \vec{\delta} p^{'} = \vec{\delta} p - D_V^2(\vec{q})\delta \vec{q} \Delta t
          \end{array}
          \right.
              \label{eq:b13}
     \end{split}
  \end{equation}

Following \cite{Skokos:2009}, the corrector operator $C$ yields the following resolvent operator
 \begin{equation}
    e^{\Delta t L_{C}}:\left\{
                \begin{array}{ll}
                  \vec{\delta} q^{'} =\vec{\delta} q\\
              \vec{\delta} p^{'} = \vec{\delta} p - D_C^2(\vec{q})\delta \vec{q} \Delta t.
          \end{array}
          \right.
              \label{eq:b17}
  \end{equation} 
Here $D_C^2(\vec{q}) =\frac{\partial^2 C }{ \partial q_n \partial q_m}$ is the Hessian.
\begin{widetext}

From Eq.~\ref{eq:b17}, we get  
  \begin{equation}
                  e^{\Delta t L_C}:\left\{
          \begin{array}{ll}
              \delta q_n^{'} = \delta q_n\\
                \delta p_n^{'} =\delta p_n -  E_J^2 \Big\{ \big[ 2 \cos(q_{n-2} - q_{n-1})  \cos(q_{n} - q_{n-1})\big] \delta q_{n-2}\\
 \qquad +\big[- 2 \cos(q_{n-1} -2 q_n+ q_{n+1})- 4\cos(2 (q_{n}-q_{n-1})) -2 \cos(q_{n-2} - 2 q_{n-1}+ q_{n})\big] \delta q_{n-1}\\
\qquad+\big[ 4 \cos(2 (q_{n+1}-q_n))+ 4 \cos(q_{n-1} -2 q_n+ q_{n+1}) +4 \cos(2 (q_{n-1}-q_n)) - 2 \sin(q_{n+2}-q_{n+1}) \sin(q_{n} \\
 \qquad- q_{n+1})-2 \sin(q_{n-2} - q_{n-1})  \sin(q_{n} - q_{n-1}) \big] \delta q_{n}\\
\qquad +\big[ - 4 \cos(2 (q_{n+1}-q_n))- 2 \cos(q_{n-1} -2 q_n+ q_{n+1})  - 2 \cos(q_{n+2}-2 q_{n+1}+q_{n})  \big] \delta q_{n+1}\\
\qquad +\big[ 2 \cos(q_{n+2}-q_{n+1}) \cos(q_{n} - q_{n+1}) \big] \delta q_{n+2}
\Big\}  \Delta t
          \end{array}
          \right.
              \label{eq:b21}
  \end{equation}
  \end{widetext}
%-------------------------------

  \section{Numerical Simulation} \label{ap3}
  We simulate Eqs. 2 in the main text with periodic boundary conditions $p_1 = p_{N+1}$ and $q_{1} = q_{N+1}$ and time step $\Delta t=0.1$. In the simulation, the relative energy error $\Delta E = |\frac{E(t)-E(0)}{E(0)}|$ is kept lower than $10^{-4}$.
The initial conditions follow by fixing the positions to zero $q_n=0$ and by choosing the moments $p_n$ according Maxwell's distribution.
The total angular momentum $L=\sum_{n=1}^N p_n$ is set zero by a proper shift of all momenta $p_n-L/N$.
Finally, we rescale $|p_n| \rightarrow a |p_n|$ to precisely hit the desired energy (density).  
%-------------------------
                 \begin{figure}[H] 
    \includegraphics[width=8.0cm,height=8cm]{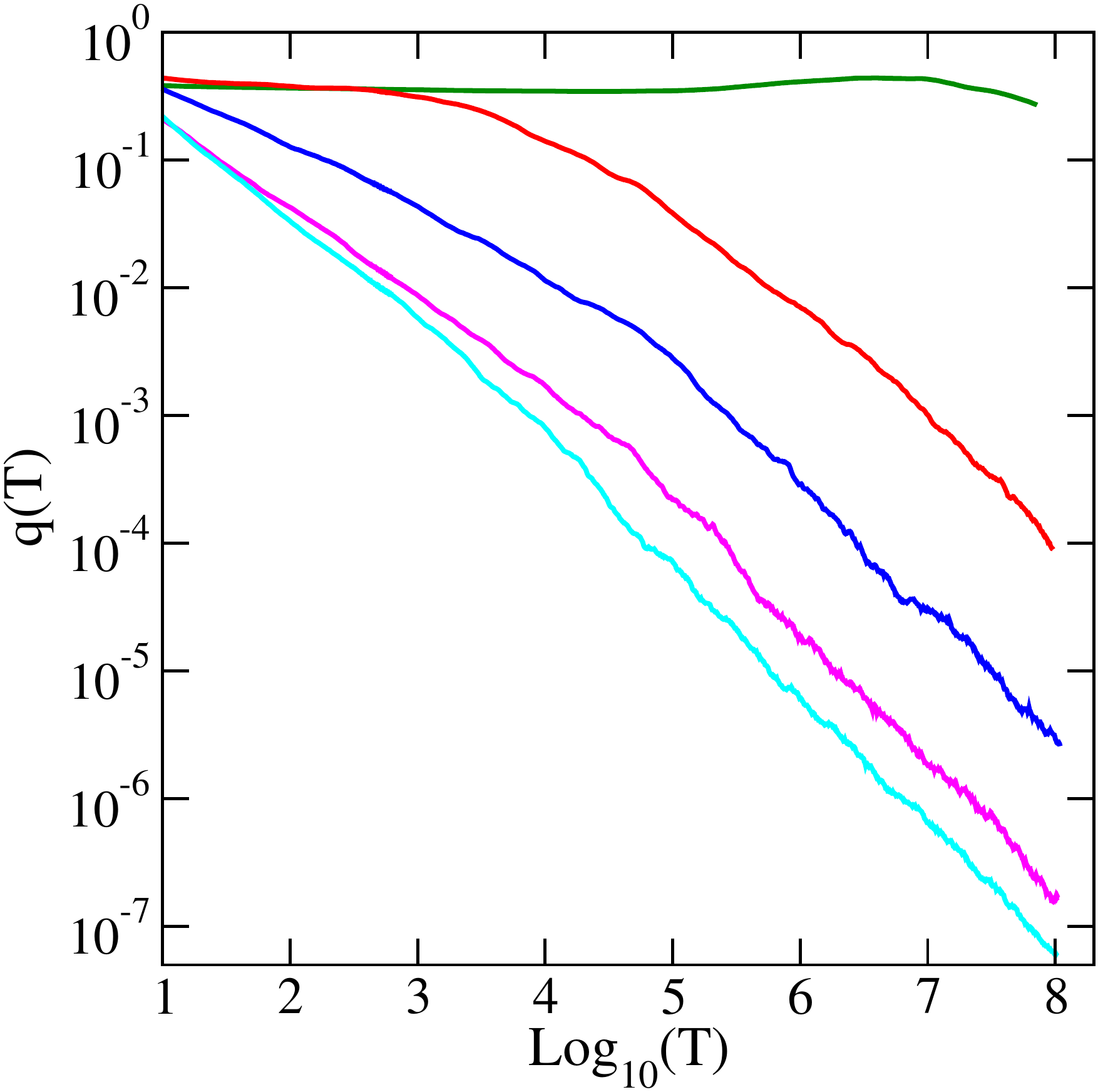}\\
    \vspace{-20.0pt}
  \caption{\label{fig_s1}{\footnotesize (Color online) Fluctuation index $q$ for fixed the energy density $h=1$ with $R=12$. 
From top to bottom:  $E_J=0.1$ (green),  $E_J=0.5$ (red),  $E_J=1.0$ (blue), $E_J=2.0$ (magenta) and  $E_J=3.0$ (cyan).}}
   \end{figure}
     %--------------------------
                   \begin{figure}[H] 
  \includegraphics[scale=1,width=8.5cm]{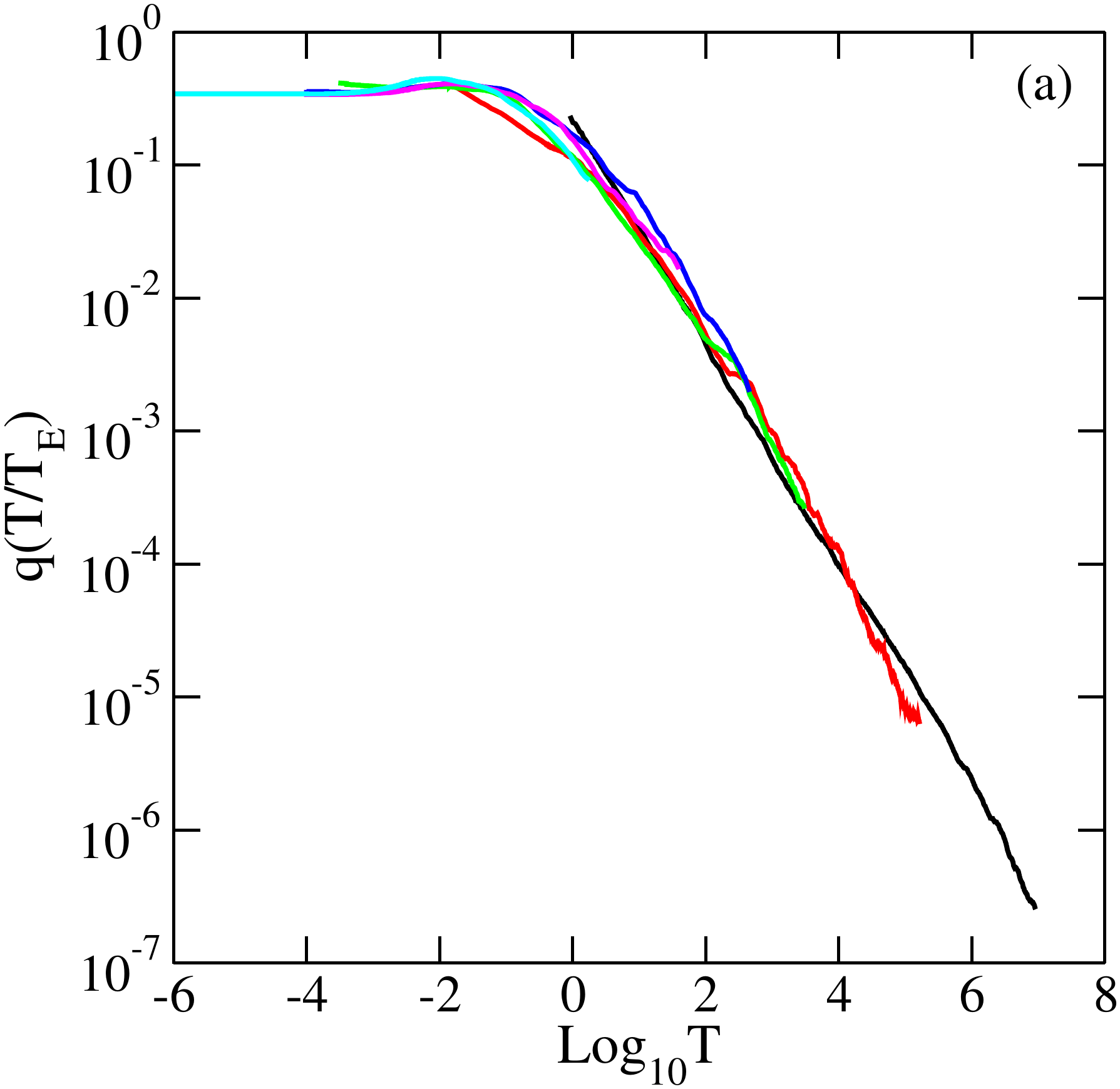}\ 
    \includegraphics[scale=1,width=8.5cm]{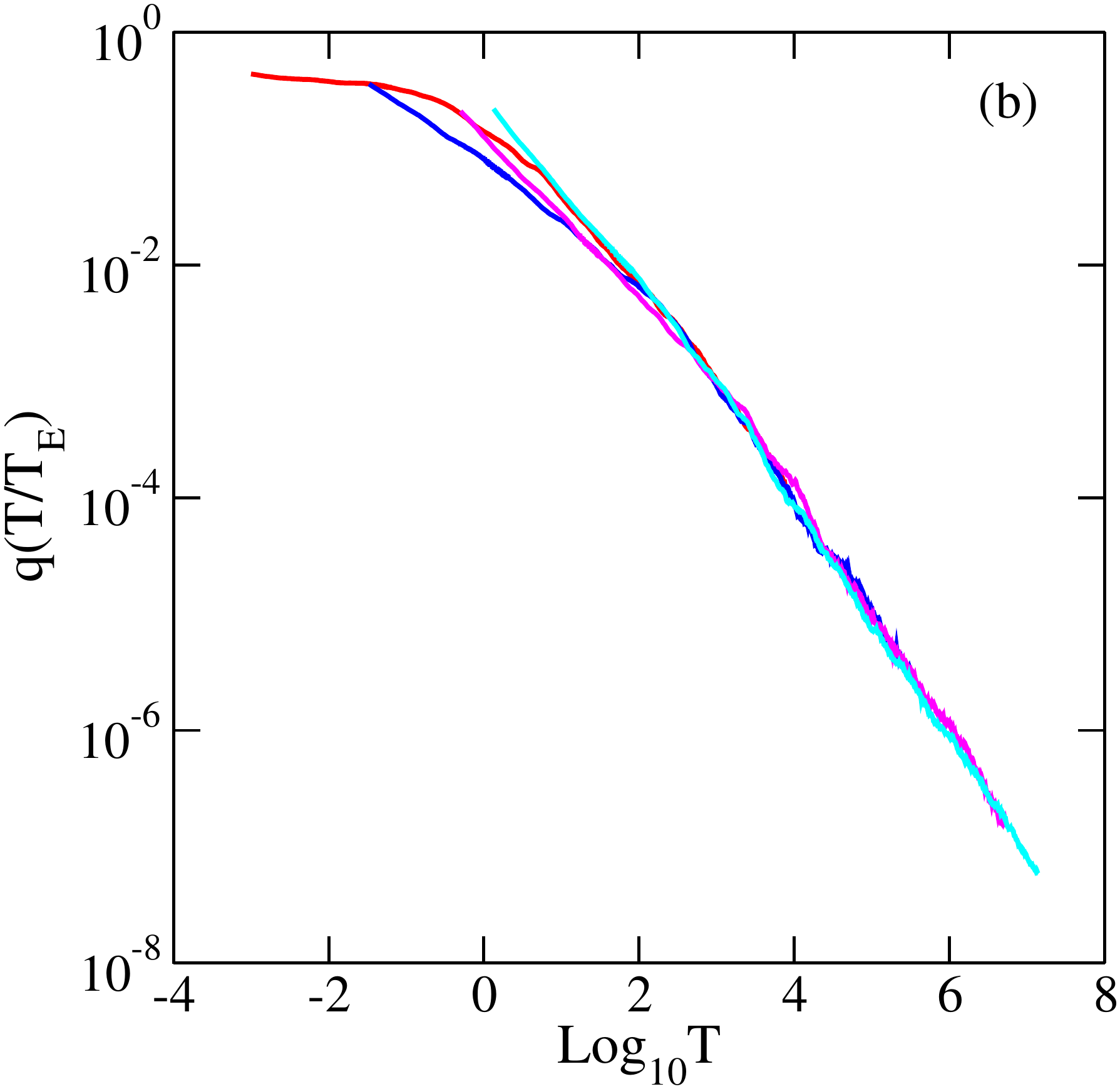} 
  \caption{\label{fig_s2}{\footnotesize (Color online) a) $q(T/T_E)$ for fixed $E_J=1$ with energy densities 0.1 (black) 1.2 (red), 2.4 (green), 3.8 (blue), 5.4 (magenta), and 8.5 (cyan) (corresponding to Fig.~1 in the main body); b) $q(T/T_E)$ for fixed energy density $h=1$ with $E_J=0.5$, (red),  $E_J=1.0$, (blue),   $E_J=2.0$, (magenta) and  $E_J=3.0$, (cyan) (corresponding to Fig.~\ref{fig_s1}).}}
   \end{figure} 
      %-----------------------------------------------
\section{Finite time average for $h=1$}\label{ap4}
Fig.~\ref{fig_s1} shows the index $q(T)$  for fixed energy $h=1$ with varying coupling strengths, $E_J$. It is similar 
to Fig.1 from the main text for fixed $E_J$ and varying $h$.
       %-------------------------------------------------------
\section{Evaluation of the ergodization time}\label{ap6} 
We rescale and fit the fluctuation index $q(T)$ shown in Figs. 1 (main body) and \ref{fig_s1} (in the supplement). 
We choose a parameter set
with a clearly observed asymptotic $q(T) \approx T_E/T$ dependence, and fit this dependence to obtain $T_E$. We then  rescale the variable $T \rightarrow xT$ for all other lines to obtain the best overlap with the initially chosen line as shown in Fig.~\ref{fig_s2}. The scaling parameters $x$ are then used to
compute the corresponding ergodization times.
      %-----------------------------------------------------------------------
      
\section{Estimate of the ergodization time $T_E$ }\label{sec:q_1/T}

In the main text, we defined  the {\it ergodization time} $T_E$ as the prefactor of the $1/T$ decay of the fluctuation index: $q(T\ll T_E) = q(0)$ and $q(T\gg T_E) \sim T_E/T$. 
We estimate this prefactor %of the $1/T$ decay of $q$ 
by approximating the time-dynamics of the observables $k_n(t)$ with  telegraphic random  process \cite{richard1972feynman,Chakravarty1985photoinduced,kac1963van}.
\begin{equation}
k_n (t) \approx % \ \ \longmapsto\ \ 
\left\{
\begin{array}{ll}
k + \alpha \qquad   \text{if}  &k_n(t) > k\\
k - \beta    \qquad  \text{if}  &k_n(t) < k
\end{array}
\right.
\label{eq:signal}
\end{equation}
with real constant $\alpha,\beta$ 
(for example see Fig.~4(a) of the main text, where $\alpha = 1 - k $ and $\beta = k $).
This recast the finite time average $\overline{k}_{n,T} = \frac{1}{T} \int_0^{T} k_n(t)dt$ of $k_n$ to
\begin{equation}
\begin{split}
\overline{k}_{n,T} 
&\approx  \frac{k + \alpha}{T}   \sum_{i=1}^{M_n^+ } \tau_{n}^+(i) + \frac{ k - \beta}{T} \sum_{i=1}^{M_n^- } \tau_{n}^-(i) \\
&\equiv k + \frac{1}{T}\left[  \alpha S_n^+ - \beta S_n^-\right]
\end{split}
\label{eq:t_ave_1_app}
\end{equation}
since  $T = S_n^+ + S_n^-$.  Here $M_n^\pm$ denote the number of excursions $\tau^\pm$ within a time interval $[0,T]$.
As  $\mu_k(T \rightarrow \infty) =k$ and  $\sigma_k(T \rightarrow \infty) =0$ (see main text), 
we  focus only on the variance $\sigma_k(T)$ to estimate the dependence on  $T$ of the fluctuation index  $q(T)=\frac{\sigma_k^2(T) }{\mu_k^2(T)}$. 
%Since $P_+$ dominates over $P_-$, w
We first rewrite Eq.(\ref{eq:t_ave_1_app}) in terms of $S_n^+$ only by adding and subtracting $\beta S_n^+ / T$ 
\begin{equation}
\begin{split}
\overline{k}_{n,T} 
&\approx  \frac{ \alpha + \beta }{T}  S_n^+ + k - \beta
\end{split}
\label{eq:t_ave_2_app}
\end{equation}
By recalling that  $\sigma_{k + B}^2(T) = \sigma_{k}^2(T) $ and $\sigma_{B k}^2(T) = B^2 \sigma_{k}^2(T) $ for any constant $B$, and by dropping the constant factor $(\alpha + \beta)$ it follows that  $\sigma_k^2(T) $  scales as
\begin{equation}
\sigma_{ k}^2(T) \sim   \frac{ 1}{T^2}  \sigma_{ S_n^+}^2(T)
\label{eq:var_Bk1}
\end{equation}
As the excursion times $\tau^+$ are considered independent variables identically distributed, it follows that 
\begin{equation}
 \sigma_{ S_n^+}^2 (T) = M_n^+(T)  \sigma_{ \tau}^2
\label{eq:var_tau}
\end{equation}
with $\sigma_{ \tau}^2$ the variance of $\tau^+$. 
For $T\gg \mu_{\tau}$ with $\mu_\tau$ the first moment of $\tau^+$,  the number of events $M_n^+$ grows as $M_n^+ \sim T / \mu_{\tau}$. 
This yields 
\begin{equation}
\sigma_{ k}^2(T) \sim   \frac{\sigma^2_{\tau}}{\mu_{\tau}} \frac{ 1}{T} 
\label{eq:var_Bk2}
\end{equation}
and ultimately to Eq.~(4) of the main text.

            %-----------------------------------------------------------------------
\section{Lyapunov exponent computation}\label{ap5}
%
  %----------------------------------
                   \begin{figure}[!htbp] 
  \includegraphics[scale=1,width=8.5cm]{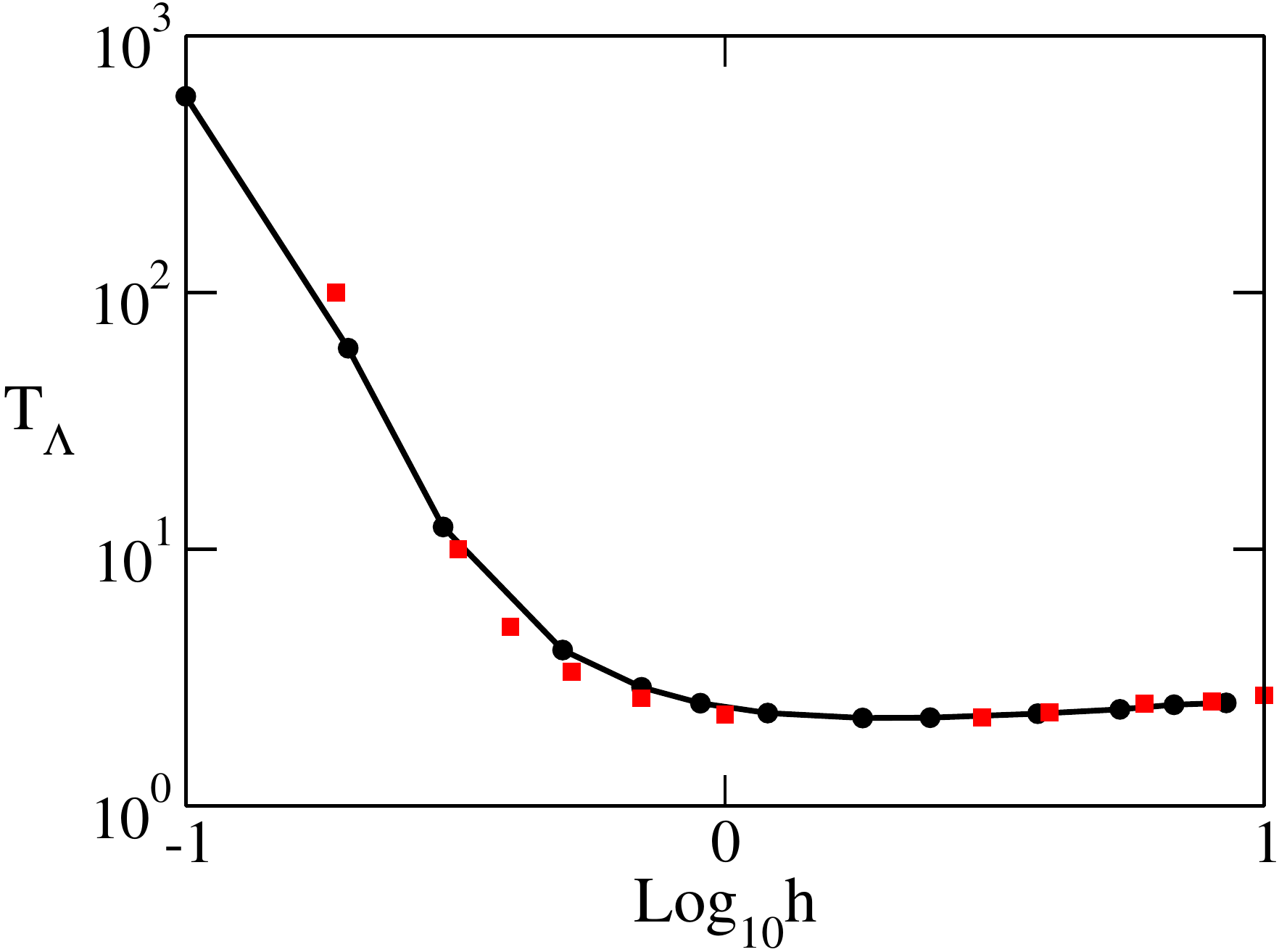} 
  \caption{\label{fig_s3}{\footnotesize (Color online) Lyapunov time for fixed $E_J=1$. The black circles represent our numerical results, and the red squares represent the analytical results from \cite{Casetti:1996}. The black line guides the eye. }}
   \end{figure}  
    %-------------------------------
We compute the largest Lyapunov exponent $\Lambda$ by numerically solving the variational equations 
\begin{equation}
\dot{ w}(t) =\big[  J_{2N} \cdot D_H^2 ( x(t) ) \big] \cdot w(t)
\label{eq:var_eq}
\end{equation}
for a small amplitude deviation $ w(t) =  (\gamma q(t) , \gamma p(t))$ coordinates.
The largest Lyapunov exponent $\Lambda$ follows
\begin{equation}
\Lambda= \lim_{t\rightarrow \infty}  \frac{1}{t}\ln \frac{\| w(t)\|}{\| w(0)\|}.
\label{eq:Lyap_1}
\end{equation} 
       In Fig.~\ref{fig_s3} we show the Lyapunov time $T_\Lambda=1/\Lambda$ versus  energy densities $h$ for given $E_J=1$. It matches well with the analytical results obtained in \cite{Casetti:1996}.

  %
%  \bibliographystyle{apsrev4}
%\let\itshape\upshape
%\normalem
%\bibliography{reference1}
 \end{document}